%% file: sample-manuscript.tex
\documentclass[acmsmall,manuscript,screen]{acmart}

\input{chat_bubbles_CJ}

\AtBeginDocument{%
  \providecommand\BibTeX{{%
    \normalfont B\kern-0.5em{\scshape i\kern-0.25em b}\kern-0.8em\TeX}}}

\setcopyright{acmlicensed}
\copyrightyear{2024}
\acmYear{2024}
\acmDOI{XXXXXXX.XXXXXXX}

\begin{document}

\title[From Keyboard to Chatbot]{From Keyboard to Chatbot: An AI-powered Integration Platform with Large-Language Models for Teaching Computational Thinking for Young Children}

\author{Changjae Lee}
\email{changjae@buffalo.edu}
\author{Jinjun Xiong}
\email{jinjun@buffalo.edu}
\affiliation{%
  \institution{University at Buffalo}
  \city{Buffalo}
  \country{USA}
}

\renewcommand{\shortauthors}{Lee and Xiong}


\input{abstract_JX}




\maketitle

\input{intro_JX}

\input{spark_user_interaction_CJ}

\input{spark_system_design_CJ}

\input{spark_chatbot_design_CJ}

\input{spark_robot_agent_CJ}

\input{spark_spark_programming_language_CJ}

\input{spark_llm_for_nlp_to_spl_CJ}

\input{conclusion_CJ}


\bibliographystyle{ACM-Reference-Format}
\bibliography{sample-base}










\end{document}

%% file: chat_bubbles_CJ.tex
\usepackage{tikz}
\usetikzlibrary{calc}

\makeatletter
\newcommand*\negcircled[1]{\tikz[baseline=(char.base)]{
    \node[shape=circle, draw, fill=black, text=white, inner sep=0.3pt, 
        minimum height={\f@size*0.8},] (char) {\vphantom{WAH1g}#1};}}
\makeatother

\usepackage{multicol}

\usepackage{transparent}

\usepackage{subcaption}

\usepackage{environ}
\usepackage{varwidth}
\usepackage[most]{tcolorbox}
\definecolor{bubblegreen}{RGB}{103,184,104}
\definecolor{bubblegray}{RGB}{241,240,240}

\tcbset{
  commonoptions/.style={
    enhanced,
    enhanced jigsaw,
    fonttitle=\bfseries,
    boxed title style={opacityback=0, opacityframe=0},
    sharp corners,
    arc=4mm,
    opacityframe=0,
    after skip=0mm,
    },
  }

\newtcolorbox{rightbubbles}[4][]{
    commonoptions,
    attach boxed title to top right={yshift=-7mm, xshift=-2mm},
    top=6mm,
    right skip=1.5cm,
    finish={\node[anchor=center, yshift=-7mm,xshift=7mm] at (frame.north east) {\roundperson{#4}};},
    title={#3:},
    rounded corners=north,
    rounded corners=southwest,
    colback=#2,
    colupper=white,
    before skip=1mm,
    #1}

\newtcolorbox{rightbubblesfinal}[2][]{
    commonoptions,
    right skip=1.5cm,
    rounded corners=northwest,
    rounded corners=south,
    colback=#2,
    colupper=white,
    before skip=0mm,
    #1}

\NewEnviron{rightbubbleslittle}[2]{\vspace{1mm}%
\tcbox[commonoptions,
    finish={\node[anchor=center, yshift=-7mm,xshift=7mm] at (frame.north east) {\roundperson{#2}};},
    rounded corners,
    colback=#1,
    colupper=white,
    before=\hfill,
    right skip=1.5cm,
    varwidth upper]
    {\BODY}
}
\newtcolorbox{leftbubbles}[4][]{
    commonoptions,
    attach boxed title to top left={yshift=-7mm, xshift=2mm},
    top=6mm,
    left skip=1.5cm,
    finish={\node[anchor=center, yshift=-7mm,xshift=-7mm] at (frame.north west) {\roundperson{#4}};},
    title={#3:},
    rounded corners=north,
    rounded corners=southeast,
    colback=#2,
    colupper=black,
    coltitle=black,
    before skip=1mm,
    #1}

\newtcolorbox{leftbubblesfinal}[2][]{
    commonoptions,
    left skip=1.5cm,
    rounded corners=northeast,
    rounded corners=south,
    colback=#2,
    colupper=black,
    before skip=0mm,
    #1}

\NewEnviron{leftbubbleslittle}[2]{\vspace{1mm}%
\begin{flushleft}
\tcbox[commonoptions,
    rounded corners,
   finish={\node[anchor=center, yshift=-7mm,xshift=-7mm] at (frame.north west) {\roundperson{#2}};},
    colback=#1,
    colupper=black,
    left skip=1.5cm,
    varwidth upper]
    {\BODY}
\end{flushleft}
}

\newcommand{\roundperson}[1]{
\begin{tikzpicture}\clip (0,0) circle (0.7cm) node {\includegraphics[width=1.4cm]{#1}};\end{tikzpicture}
}

\newtcolorbox{mybox}[1]{colback=blue!5!white,colframe=blue!75!black,fonttitle=\bfseries,title=#1, width=\textwidth}

%% file: abstract_JX.tex
\begin{abstract}
Teaching programming in early childhood (4-9) to enhance computational thinking has gained popularity in the recent movement of computer science for all. 
However, current practices ignore some fundamental issues resulting from young children's developmental readiness, such as the sustained capability to keyboarding, the decomposition of complex tasks to small tasks, the need for intuitive mapping from abstract programming to tangible outcomes, and the limited amount of screen time exposure. To address
these issues in this paper, we present a novel methodology with an AI-powered integration platform 
to effectively teach computational thinking for young children. The system features a hybrid pedagogy that supports both the top-down and bottom-up approach for teaching computational thinking. Young children can describe their desired task in natural language, while the system can respond with an easy-to-understand program consisting of the right level of 
decomposed sub-tasks. A tangible robot  can immediately execute the decomposed program and demonstrate the program's outcomes to young children. The system is equipped with an intelligent chatbot that can interact with young children through natural languages, and children can speak to the chatbot to complete all the needed programming tasks, while
the chatbot orchestrates the execution of the program onto the robot. This would completely eliminates the need of keyboards for young children to program.
At the heart of the system is a domain-specific Spark programming language that provides a right abstraction for the underlying computing hardware (i.e., robot in this case) and acts as a bridge between young children's natural language interface and the necessary programming constructs for executing the program. Young children will learn the concepts of programming and computational thinking by reading (or listening to) the programs as expressed by the Spark programming language.
Modern large-language models (LLMs) are essential to provide the capability for our system  to  support the semantic decomposition
    of high-level programming tasks as expressed in children's natural language 
    to a set of low level tasks as expressed in Spark programming language. 
By developing such a system, 
 we aim to make the concept of computational thinking  more accessible to young children, fostering a natural understanding of programming concepts without the need of explicit programming skills. Through the interactive experience provided by the robotic agent, our system seeks to engage children in an effective manner, contributing to the field of educational technology for early childhood computer science education. 
\end{abstract}

%% file: intro_JX.tex
\section{Introduction}

As computing technology becomes more prevalent in people's daily life,  
engaging young children early to be compute-literate has received increasing
attention from both parents and educators alike~\cite{blikstein2018pre, price2018evaluation}. 
Computational thinking~\cite{wing2006computational, proctor2023computational} is one of the
most prevalent conceptual frameworks for teaching young children about computing. It emphasizes the importance of  learning some fundamental analytical skills 
that are universal for any problem solving contexts, skills such as decomposition of a large complex task into smaller, 
manageable tasks recursively, and organizing the solution
process through basic computing constructs 
such as sequencing, conditional branching, 
and repeated loops. Although computational thinking is different from or even beyond computer science (CS),
computational thinking has 
catalyzed a slew of policy changes around the notion of \textit{CS for All}, thanks to, in part,
coalition efforts of big organizations like Association for Computing Machinery (ACM), code.org etc. 
Most education activities nowadays teach computational thinking  
through teaching programming as part of computer science education~\cite{BERS2014145, SU2023100122, Lee2023-xw, designing_scratchjr, scratchjr_sys_lit_review, kibo_robot_demo, survey_of_comp_kits, s23239335, Papadakis2020gy},
and being able to
program has become a fundamental competency for children receiving either primary or secondary education.

This way of teaching computational thinking may be effective for adults or big children (say 10 or above), but for young children (say 4 to 9), it may be less effective due to a number of developmental challenges
associated with young children in learning to program. 
(1) First, keyboarding and typing is almost a prerequisite skill for learning
programming (no matter what programming languages or programming interfaces to use), but
according to the United States Common Core State Standards~\cite{ccss}, 
most children in the 4th Grade are capable of
typing set about 300 words (i.e., 1 page of text) in
a session.
This is clearly beyond the normal capabilities
of young children when they are required to program through keyboarding. Strategies of having adults to sit by young children to help
with typing or give click-by-click instructions will only pose counter-intuitive challenges for
children's learning~\cite{beals2006robotic}.
(2) Second, most programming education  starts 
with teaching the basic low-level instructions (such
as variables, assignment, branching conditions, and repeated loops),
and many of them do not have a visible outcome by themselves. This lack of direct association between instructions
and their results is a major factor that hinders children's intuitive learning~\cite{designing_scratchjr}.
(3) Third, existing pedagogy of teaching young children to program mostly adopts a bottom-up computer science programming approach, i.e., first to learn the meaning of low-level instructions, and then to build the more
complex functions bottom-up level by levels. In other words, this requires learners knowing 
\textit{a priori} how to decompose
a high-level task into several levels of low-level tasks in the form of functions (or sub-components), a skill most young children may have not learned yet.
This seemingly contradictory requirements for young children often induce distraction, frustration, and disappointment
for young learners, further jeopardizing their learning effectiveness~\cite{designing_scratchjr}.
(4) Fourthly, the amount of screen time required to code and debug a reasonably complex program may be too
much for young children, and this is often cited as one of the main concerns from educators in 
adopting many modern programming teaching tools for young children~\cite{kibo_robot_demo}.

There are a number of recent attempts trying to address
some of these issues. For example, ScratchJr~\cite{designing_scratchjr} simplifies the required low-level constructs by providing
a more restricted
set of programming constructs (and it even removed explicit conditional branching blocks) than its predecessor 
Scratch~\cite{resnick2009scratch} while leveraging the same screen-based graphical programming frameworks.  It only alleviates some of
the programming complexity issues at the cost of sacrificing functionality, 
yet it still suffers the same bottom-up mapping dilemma issues as well keyboarding and screen-time challenges.
To address the lack of mapping from low-level instructions to their results, 
KIBO~\cite{kibo_robot_demo} proposes a novel  programming interface that is
developmentally appropriate. KIBO consists of a robot and a a set of
tangible wooden programming blocks. Children learn to program by interlocking 
those tangible wooden blocks, which in turn determine how the robot should move and react. KIBO saves young children's screen-time for programming
and builds the mapping between programming logic and results in a tangible and straightforward way.
But it still follows a bottom-up learning approach and requires young children knowing about
decomposition in order to build the program. The complexity of the program is also limited due to the
limited number of tangible wooden blocks.
A very recent academic prototype, TurtleTalk~\cite{tutletalk_vui}, removes the need of keyboarding by introducing a Voice User Interface (VUI). However, it is a simple wrapper of the traditional block-based graphic programming
interface like ScratchJr, hence it suffers similar drawbacks of ScratchJr in terms of its complexity
of programming. The entire interface still requires a lot of screen time for interaction.

These highlighted problems across various programming education platforms
lead us to a fundamental research question, i.e., 
``What is a natural and intuitive approach to teach computational thinking for young childhood?''
We assume that young children in this age group (4-9) would have a reasonable language capability to articulate
any tasks they desire to program, but they still lack a sustained ability for keyboarding, and their capabilities for complex decomposition of
high-level tasks to low-level program constructs
are limited (and to be developed).

In response to these challenges, we propose a novel methodology with an integrated platform, Spark, for teaching young children
about computational thinking in this paper. Different from existing solutions, Spark leverages 
the various latest development of modern artificial intelligence technologies, and it possesses the 
following characteristics:
\begin{itemize}
    \item A hybrid pedagogy that supports both the top-down and bottom-up approach for teaching computational thinking where young children can describe their desired task in natural language, be it high-level tasks or low-level tasks, while the system can respond with an easy-to-understand program consisting of the right level of decomposed sub-tasks. 
    \item A robot that can immediately execute the decomposed program and demonstrate the program's outcomes to young children in a direct and tangible fashion. The mapping of the programs to the
    robot's basic functionalities are transparent to the young children and handled by our integrated platform.
    \item An intelligent chatbot that provides a natural language-based interface to support young children's natural interaction with the system. Children can talk to the chatbot to realize the programming to accomplish a desired high level task. Through the same natural language interface, children can revise the program, 
    build reusable function modules, and determine
    the execution of the program. The chatbot orchestrates the execution of the program on the robot, and it completely eliminates the need for young children to type the program through keyboards (though we still provides a typing interface for flexibility).
    \item A domain-specific Spark programming language that provides a right abstraction for the underlying robot and acts as a bridge between young children's natural language for programming and the necessary programming constructs for executing the program. Young children will learn the concepts of programming and computational thinking by reading (or listening to) the programs as expressed by the Spark programming language.
    \item A large-language model (LLM) based decomposition methodology that supports the semantic decomposition
    of high-level tasks as expressed in children's natural language 
    to a set of low level tasks as expressed in Spark programming language. 
    \item An integrated programming platform that connects all the required pieces to provide an easy-to-use,
    developmentally appropriate interface for young children to interact with (and to program) the robot while
    learning the relevant computational thinking concepts, including decomposition, sequencing, branching,
    and repeated loops.
\end{itemize}

With such an integrated platform, a lot of interesting use cases can be built around it to support
various research investigations, such as how young children develop the computational thinking skills
when interacting with a robot and program it through a natural language interface, a topic we plan to do in our future work.

The rest of the paper is organized as follows. Section~\ref{sec_user_interaction} shows a number of
typical work flows of the proposed Spark system in helping young children to learn the concepts of programming and computational thinking.
Section~\ref{sec_spark_design} details the overall design of the proposed Spark system and the related
user input and output interfaces.
In Section~\ref{sec_chatbot}, we discuss the design of Spark's conversation engine (chatbot) and how it supports users' natural language interactions throughout
the entire programming process.
Section~\ref{sec_robot} describes the design of the Spark's robot engine and the Spark SDK. We will also show how this is implemented through
the Unitree Go1 robotic dog as an example.
Section~\ref{sec_spark_design} illustrates how a domain specific programming language, the Spark programming
language, is constructed to bridge the gap between
natural language and underlying hardware's programming.
In Section~\ref{LLM_for_decomposition}, we explain how LLM is leveraged to provide a semantic understanding of
users' programming intent as expressed in natural language, and how LLM helps to produce an executable program as expressed in Spark programming language.
Section \ref{sec_conclusion} concludes this paper
with discussion of our future work. We hope our methodology and our integrated platform can inspire
more systematic research into the effective education of  young children's computational thinking skills.

%% file: spark_user_interaction_CJ.tex
\section{User Interaction}
\label{sec_user_interaction}

\begin{figure}[h]
  \centering
  \includegraphics[scale=0.5]{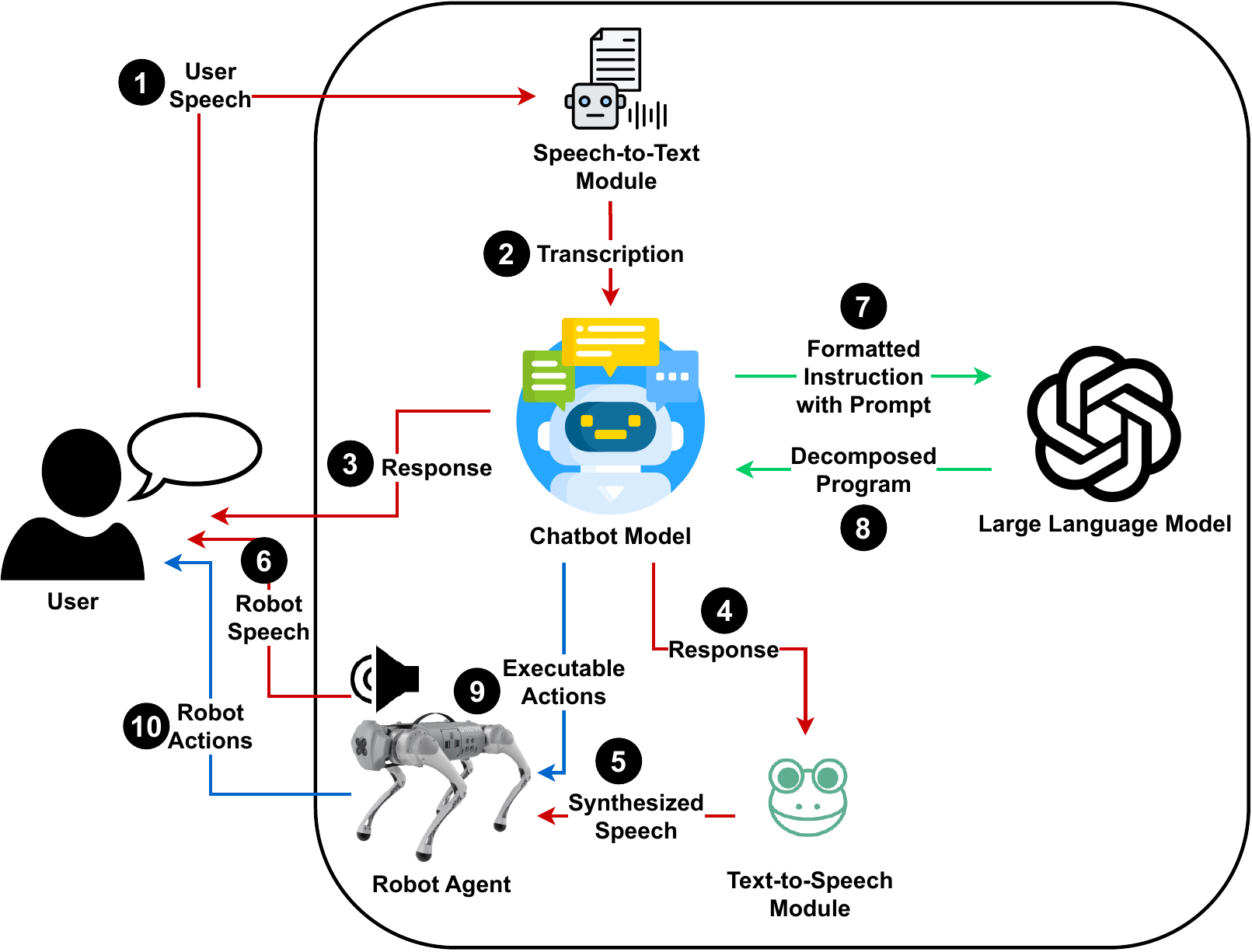}
  \caption{The high-level workflow of Spark}
  \label{workflow}
\end{figure}

\subsection{Exemplar Work Flows}
The workflow of our system in interaction with users, specifically younger children, is visually represented in Figure \ref{workflow}. The intentional exclusion of the web user interface component aims to streamline focus on the core processes. The interaction unfolds through four distinct cases, each contributing to the overall functionality of the system.

\subsubsection{General Case}
\begin{figure}[H]
\centering
\begin{rightbubbleslittle}{bubblegreen}{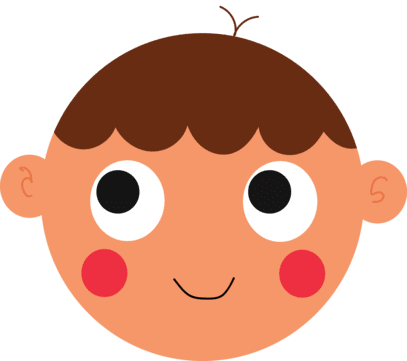}
Hey Spark!
\end{rightbubbleslittle}
\begin{leftbubbleslittle}{black!15}{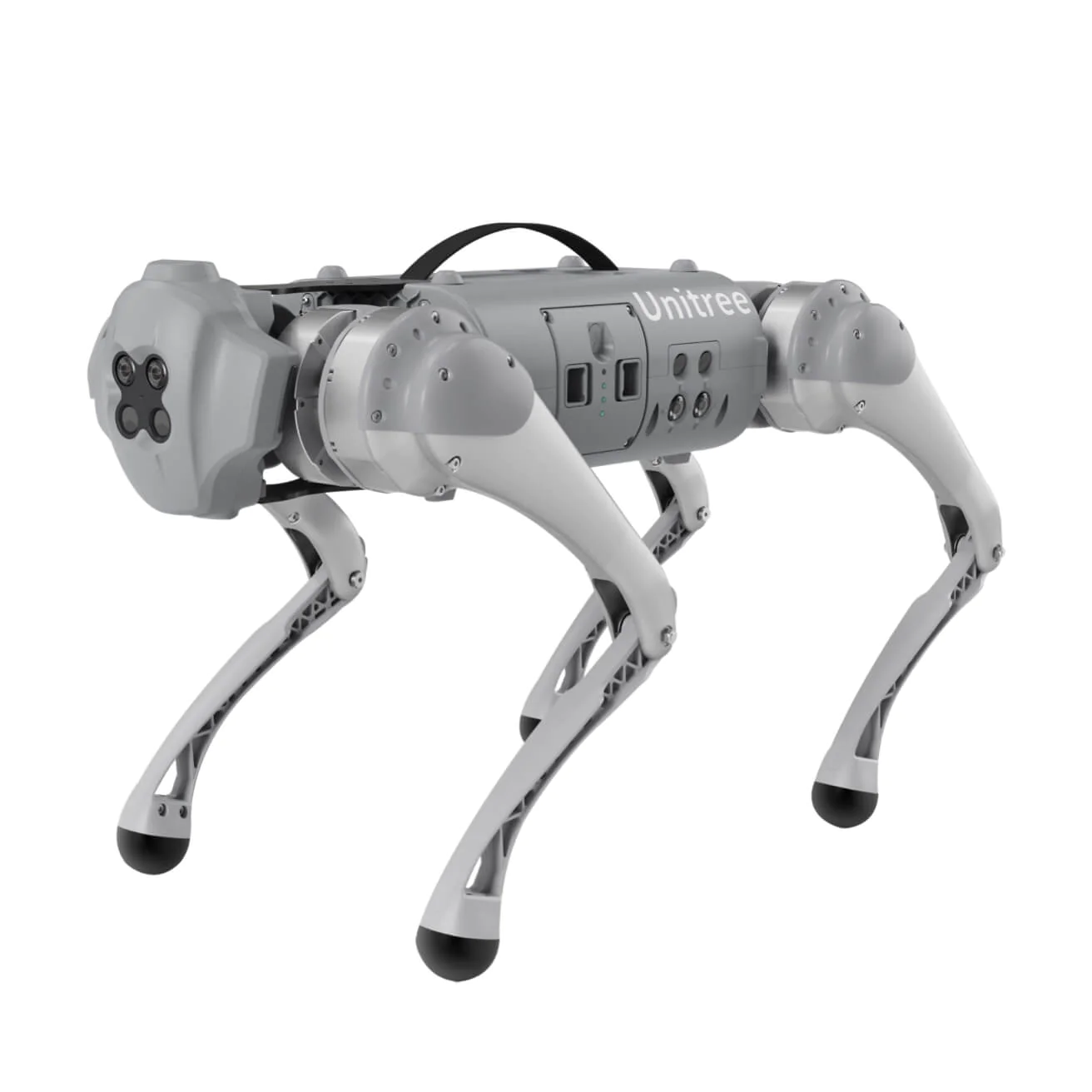}
Yes, What do you want me to do?
\end{leftbubbleslittle}
\begin{rightbubbleslittle}{bubblegreen}{child_cartoon.png}
What can you do?
\end{rightbubbleslittle}
\begin{leftbubbleslittle}{black!15}{unitree_go1_image.png}
I can do the following easy actions: STAND\_DOWN, STAND\_UP, TILT\_LEFT\_SHOULDER, TILT\_RIGHT\_SHOULDER, TILT\_HEAD\_UP, TILT\_HEAD\_DOWN, TILT\_HEAD\_LEFT, TILT\_HEAD\_RIGHT, MOVE\_FORWARD, MOVE\_LEFT, MOVE\_RIGHT, TURN\_LEFT, TURN\_RIGHT, SPIN\_JUMP, LIFT, FIRST\_DANCE, SECOND\_DANCE, FIND \\
I can also do difficult actions by doing many easy actions.
\end{leftbubbleslittle}
\caption{An example of general case} 
\label{example_general_case}
\end{figure}

In the general case, the system's interaction is confined to users, excluding engagement with the LLM and the robotic agent. The procedural steps for this case adhere to the red-colored arrows in Figure \ref{workflow}. The detailed breakdown of this scenario is outlined as follows: \negcircled{1} Users articulate their thoughts, with the speech captured in a formatted file transmitted to the Speech-to-Text (STT) module. \negcircled{2} The STT module processes the user's speech, transcribing it into text, which is then sent to the chatbot model. \negcircled{3} The chatbot model formulates a response to the user input, presenting it in text format. \negcircled{4} The chatbot model's textual response is conveyed to the Text-to-Speech (TTS) module. \negcircled{5} The TTS module synthesizes the textual response into corresponding speech, generating a synthesized speech file. \negcircled{6} The synthesized speech file is played by the robotic agent, allowing users to listen to the system's response. A graphical depiction of this case is presented in Figure \ref{example_general_case}. 

\subsubsection{Generating Program Case}
\begin{figure}[H]
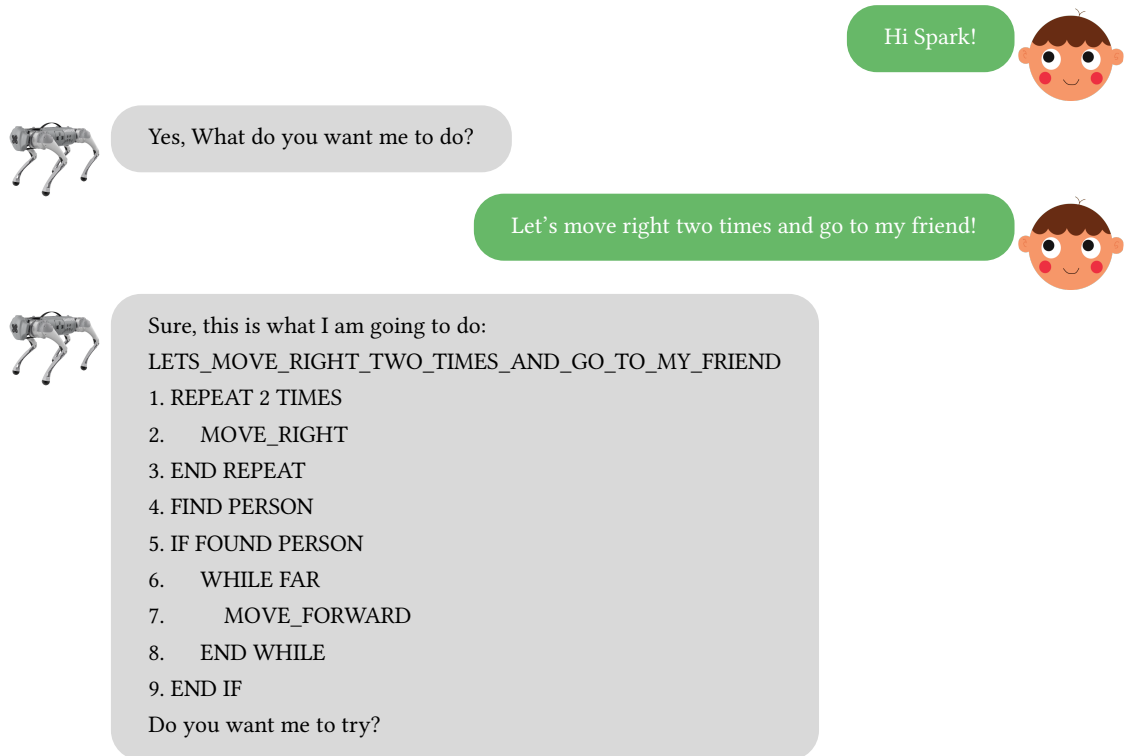

\centering
\begin{rightbubbleslittle}{bubblegreen}{child_cartoon.png}
Hi Spark!
\end{rightbubbleslittle}
\begin{leftbubbleslittle}{black!15}{unitree_go1_image.png}
Yes, What do you want me to do?
\end{leftbubbleslittle}
\begin{rightbubbleslittle}{bubblegreen}{child_cartoon.png}
Let's move right two times and go to my friend!
\end{rightbubbleslittle}
\begin{leftbubbleslittle}{black!15}{unitree_go1_image.png}
Sure, this is what I am going to do:\\
LETS\_MOVE\_RIGHT\_TWO\_TIMES\_AND\_GO\_TO\_MY\_FRIEND\\
1. REPEAT 2 TIMES\\
2. {\transparent{0}~}{\transparent{0}~}{\transparent{0}~}{\transparent{0}~}{\transparent{0}~}MOVE\_RIGHT\\
3. END REPEAT\\
4. FIND PERSON\\
5. IF FOUND PERSON\\
6. {\transparent{0}~}{\transparent{0}~}{\transparent{0}~}{\transparent{0}~}{\transparent{0}~}WHILE FAR\\
7. {\transparent{0}~}{\transparent{0}~}{\transparent{0}~}{\transparent{0}~}{\transparent{0}~}{\transparent{0}~}{\transparent{0}~}{\transparent{0}~}{\transparent{0}~}MOVE\_FORWARD\\
8. {\transparent{0}~}{\transparent{0}~}{\transparent{0}~}{\transparent{0}~}{\transparent{0}~}END WHILE\\
9. END IF\\
Do you want me to try?
\end{leftbubbleslittle}
\caption{An example of generating program case} 
\label{example_generating_program_case}
\end{figure}

The process involved in the program generation case is intricate, involving cooperation between the chatbot model, the LLM, and the robotic agent. The steps for this case, indicated by the green-colored arrows in Figure \ref{workflow} and incorporating red arrows, can be broken down as follows: \negcircled{1} Users provide specific instructions after the greeting ``hey spark'', delivered as speech to the system, which is processed by the STT module. \negcircled{2} The STT module converts the user's speech into transcription, preparing it for further processing. \negcircled{7} The chatbot model sends the user instruction along with a prompt to the LLM, requesting the corresponding semantically decomposed program written in the Spark Programming Language (SPL). \negcircled{8} The LLM generates the semantically decomposed program in SPL based on the user instruction and promptly sends it back to the chatbot model. \negcircled{3} The chatbot model displays the response, including the decomposed result from the LLM, to the user in textual format. \negcircled{4} The chatbot model's textual response is also transmitted to the TTS module. \negcircled{5} The TTS module synthesizes the given text response from the chatbot model into corresponding speech. \negcircled{6} The synthesized speech is played by the robotic agent, communicating the system's response, including the decomposed program, to the user. A visual representation of this case is provided in Figure \ref{example_generating_program_case}.

\subsubsection{Program Revision Case}
The program revision case represents a combination of the general and program generation cases, allowing younger children to modify the generated program with assistance from the LLM. Analogous to the program generation case, this process adheres to the progression outlined by the green-colored arrows. The sequential breakdown of this scenario is delineated as follows: \negcircled{1} Users articulate specific instructions following the greeting ``hey sparky'', conveyed orally to the system. \negcircled{2} The STT module converts the user's spoken input into transcribed text. \negcircled{7} The transcribed user input is then transmitted along with a prompt to the LLM. \negcircled{8} The LLM generates the semantically decomposed program in SPL based on the user instruction and promptly sends it back to the chatbot model. \negcircled{3} The resultant decomposed program is presented to the user in textual format through the chatbot model's response. \negcircled{4} The chatbot model's textual response is also transmitted to the TTS module. \negcircled{5} The TTS module transforms the provided textual response from the chatbot model into audible speech. \negcircled {6} The resulting synthesized speech is then presented by the robotic agent. 

\begin{figure}[H]
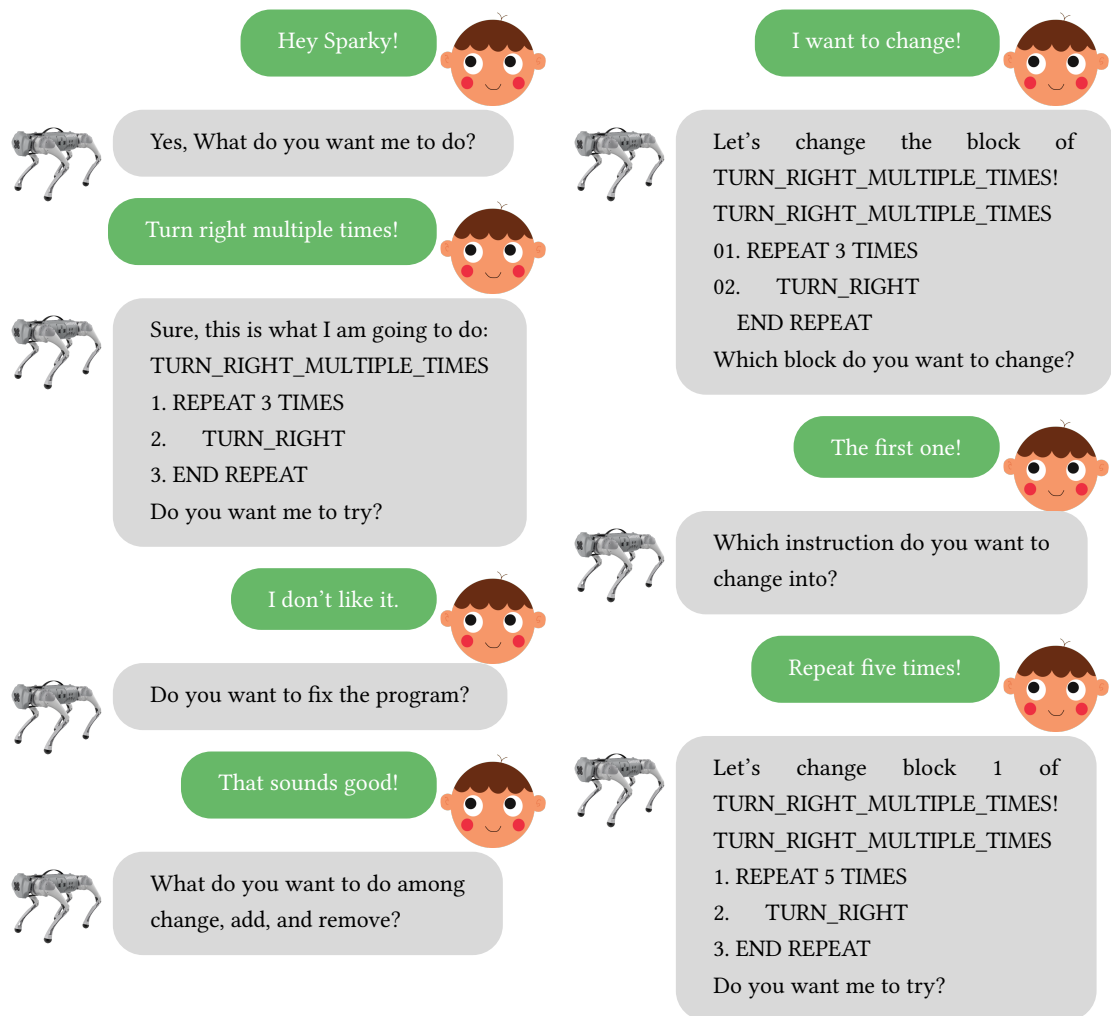

\centering
\begin{subfigure}[t]{.49\linewidth}
\begin{rightbubbleslittle}{bubblegreen}{child_cartoon.png}
Hey Sparky!
\end{rightbubbleslittle}
\begin{leftbubbleslittle}{black!15}{unitree_go1_image.png}
Yes, What do you want me to do?
\end{leftbubbleslittle}
\begin{rightbubbleslittle}{bubblegreen}{child_cartoon.png}
Turn right multiple times!
\end{rightbubbleslittle}
\begin{leftbubbleslittle}{black!15}{unitree_go1_image.png}
Sure, this is what I am going to do:\\
TURN\_RIGHT\_MULTIPLE\_TIMES\\
1. REPEAT 3 TIMES\\
2. {\transparent{0}~}{\transparent{0}~}{\transparent{0}~}{\transparent{0}~}{\transparent{0}~}TURN\_RIGHT\\
3. END REPEAT\\
Do you want me to try?
\end{leftbubbleslittle}
\begin{rightbubbleslittle}{bubblegreen}{child_cartoon.png}
I don't like it.
\end{rightbubbleslittle}
\begin{leftbubbleslittle}{black!15}{unitree_go1_image.png}
Do you want to fix the program?
\end{leftbubbleslittle}
\begin{rightbubbleslittle}{bubblegreen}{child_cartoon.png}
That sounds good!
\end{rightbubbleslittle}
\begin{leftbubbleslittle}{black!15}{unitree_go1_image.png}
What do you want to do among change, add, and remove?
\end{leftbubbleslittle}
\end{subfigure}
\begin{subfigure}[t]{.49\linewidth}
\begin{rightbubbleslittle}{bubblegreen}{child_cartoon.png}
I want to change!
\end{rightbubbleslittle}
\begin{leftbubbleslittle}{black!15}{unitree_go1_image.png}
Let's change the block of TURN\_RIGHT\_MULTIPLE\_TIMES!\\
TURN\_RIGHT\_MULTIPLE\_TIMES\\
01. REPEAT 3 TIMES\\
02. {\transparent{0}~}{\transparent{0}~}{\transparent{0}~}{\transparent{0}~}{\transparent{0}~}TURN\_RIGHT\\
{\transparent{0}~}{\transparent{0}~}{\transparent{0}~}{\transparent{0}~}END REPEAT\\
Which block do you want to change?
\end{leftbubbleslittle}
\begin{rightbubbleslittle}{bubblegreen}{child_cartoon.png}
The first one!
\end{rightbubbleslittle}
\begin{leftbubbleslittle}{black!15}{unitree_go1_image.png}
Which instruction do you want to change into?
\end{leftbubbleslittle}
\begin{rightbubbleslittle}{bubblegreen}{child_cartoon.png}
Repeat five times!
\end{rightbubbleslittle}
\begin{leftbubbleslittle}{black!15}{unitree_go1_image.png}
Let's change block 1 of TURN\_RIGHT\_MULTIPLE\_TIMES!\\
TURN\_RIGHT\_MULTIPLE\_TIMES\\
1. REPEAT 5 TIMES\\
2. {\transparent{0}~}{\transparent{0}~}{\transparent{0}~}{\transparent{0}~}{\transparent{0}~}TURN\_RIGHT\\
3. END REPEAT\\
Do you want me to try?
\end{leftbubbleslittle}
\end{subfigure}
\caption{An example of program revision case} 
\label{example_program_revision_case}
\end{figure}

Subsequently, in the event of user disapproval of the generated program, our system prompts an inquiry regarding the desire to modify the program, adhering to the designated red arrows in Figure \ref{workflow}. Upon the user's affirmation, the chatbot model proceeds to solicit the necessary information required for program revision, consistently following the prescribed red arrows. Upon receiving the user's instructions for modification (\negcircled{1} and \negcircled{2}), \negcircled{7} the chatbot model transmits this information along with a prompt to the LLM, requesting the corresponding program suitable for program revision. \negcircled{8} The LLM generates the corresponding program and returns it to the chatbot model. The ensuing steps (\negcircled{3}-\negcircled{6}) involving red elements are iteratively followed, repeating the process until the user expresses satisfaction. An illustrative example of this case is elucidated in Figure \ref{example_program_revision_case}.

\subsubsection{Program Execution Case}
The program execution case marks the concluding phase, involving user consent and the tangible execution of the generated program by the LLM and potentially revised by the users in response to the system's query ``Do you want me to try?''. Incorporating the red arrows in Figure \ref{workflow}, the procedural steps for this case are indicated by the blue-colored arrows. The detailed breakdown of this scenario is expounded as follows: \negcircled{1} Users manifest their agreement to execute the generated program, conveyed through affirmative statements such as ``Yes'' or ``Sure, please.''. \negcircled{2} The STT module converts the user's verbal expression of agreement into textual format. \negcircled{3} The chatbot model responds in text, confirming the user's agreement to execute the program. This textual confirmation is displayed to the user. \negcircled{4} The chatbot model transmits its text response to the TTS module. \negcircled{5} The TTS module transforms the provided text response into an associated speech file. \negcircled{6} The robotic agent plays the generated speech file from the TTS module, serving as an audible confirmation to the user. \negcircled{9} Internally, subsequent to dispatching the text response to the TTS module, the chatbot model translates the program scripted in the Spark Programming Language (SPL) into the corresponding Python code. Following this, it promptly executes the Python code, guiding the actions of the robotic agent. \negcircled{10} The robotic agent, operating based on the executed program, furnishes a tangible interface to the user, undertaking and completing the specified tasks. A demonstrative example of this case is depicted in Figure \ref{example_program_execution_case}.

\begin{figure}[H]
\centering
\begin{rightbubbleslittle}{bubblegreen}{child_cartoon.png}
Hello Spark!
\end{rightbubbleslittle}
\begin{leftbubbleslittle}{black!15}{unitree_go1_image.png}
Yes, What do you want me to do?
\end{leftbubbleslittle}
\begin{rightbubbleslittle}{bubblegreen}{child_cartoon.png}
Please do spin jump!
\end{rightbubbleslittle}
\begin{leftbubbleslittle}{black!15}{unitree_go1_image.png}
Sure, this is what I am going to do:\\
PLEASE\_DO\_SPIN\_JUMP\\
1. SPIN\_JUMP\\
Do you want me to try?
\end{leftbubbleslittle}
\begin{rightbubbleslittle}{bubblegreen}{child_cartoon.png}
Sure, please.
\end{rightbubbleslittle}
\begin{leftbubbleslittle}{black!15}{unitree_go1_image.png}
Let's run the program!
\end{leftbubbleslittle}
\caption{An example of program execution case} 
\label{example_program_execution_case}
\end{figure}

%% file: spark_system_design_CJ.tex
\begin{figure}[h]
  \centering
  \includegraphics[scale=0.6]{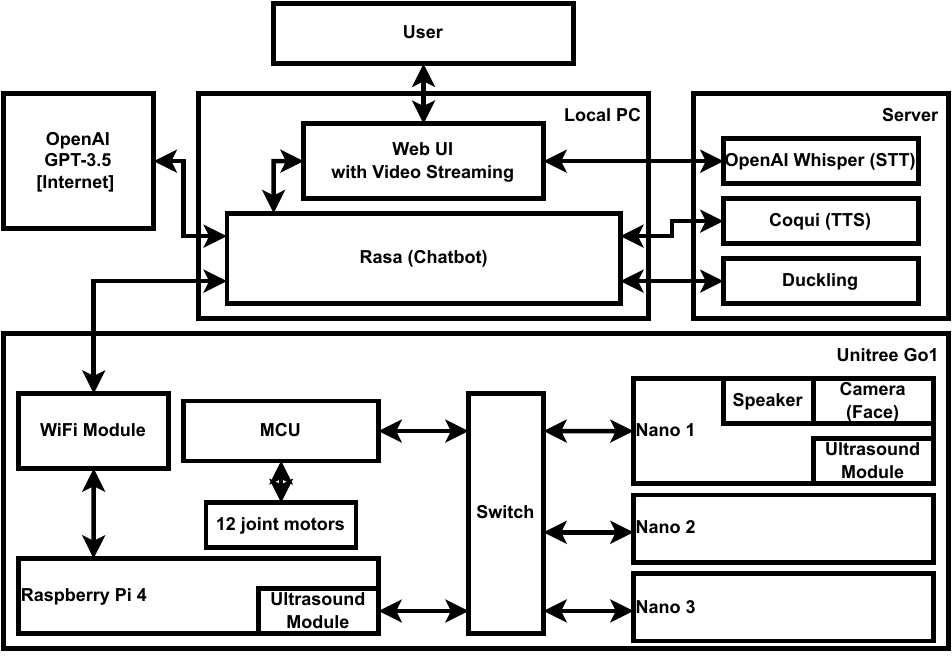}
  \caption{The architecture diagram of Spark} 
  \label{Architecture}
\end{figure}

\section{Spark Design}
\label{sec_spark_design}

Spark embodies a crafted integration that seamlessly combines three pivotal elements, each playing a distinct role in creating a comprehensive system geared towards enhancing user experience and fostering seamless interaction. The first cornerstone of Spark is the Chatbot featuring a Voice User Interface (VUI). This interactive interface facilitates natural and intuitive communication with users. Equipped with VUI, the chatbot enables dynamic and user-friendly interactions. Users can engage in conversations with the chatbot, establishing a fluid and responsive communication model that enhances the overall user experience. The second integral component is the LLM, strategically designed for task decomposition. The LLM interprets and processes user inputs, playing a pivotal role in breaking down complex programming tasks into manageable components. This functionality simplifies the programming experience, especially for younger users, by providing a structured approach to task execution. The LLM's intelligent processing contributes to a user-friendly programming environment. Completing the trio is the Tangible Robotic Companion, serving as a tangible link between programming concepts and real-world actions. This physical entity brings code to life, allowing users to witness tangible outcomes resulting from their programming efforts. The tangible robotic companion effectively bridges the gap between the virtual programming environment and the physical world, providing a hands-on experience that enhances the understanding of programming principles. Together, these three elements—Chatbot with VUI, LLM for task decomposition, and the Tangible Robotic Companion—unite to form Spark, a cohesive and synergistic integration. Spark not only facilitates programming interactions but also enriches the overall user experience by combining natural language interaction, intelligent task processing, and real-world embodiment through the tangible robotic companion. This holistic approach ensures that users, especially those new to programming, can engage with Spark seamlessly and enjoy a multifaceted learning experience. 

\subsection{User Interface}
The User Interface, stands as a pivotal medium for receiving user input within the Spark system. This interface is designed, comprising two main components that contribute to a rich and immersive user experience.

\begin{figure}[h]
  \centering
  \includegraphics[scale=0.2]{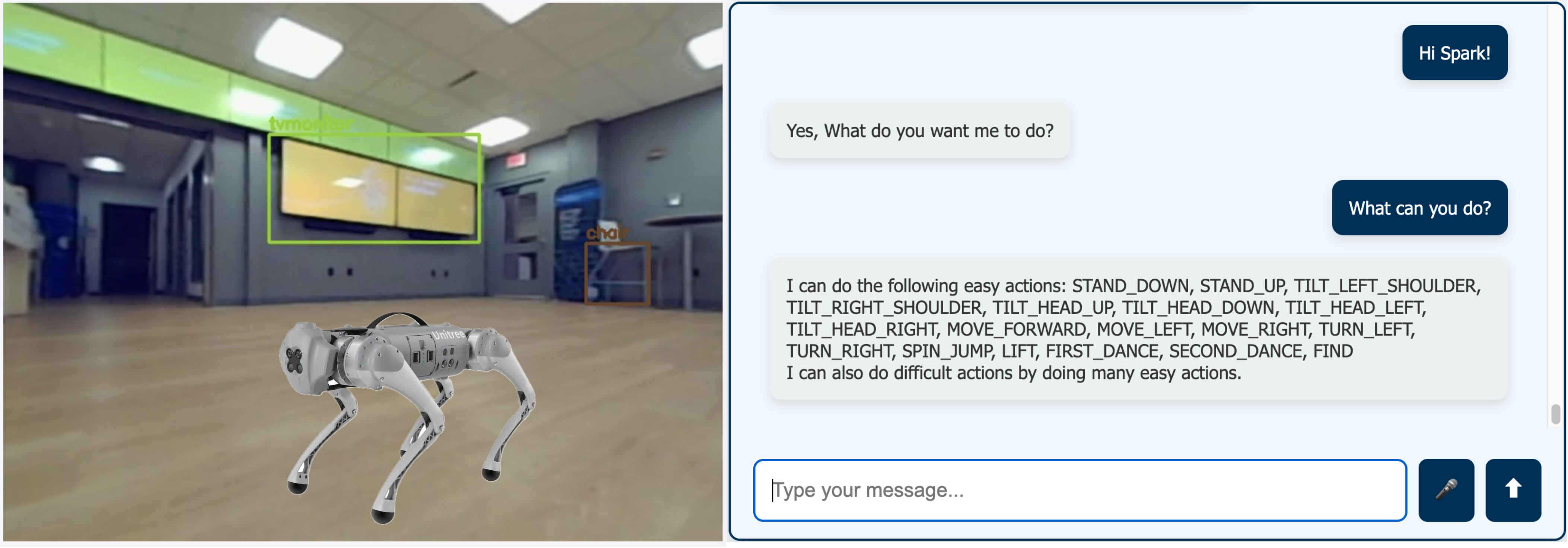}
  \caption{The user interface of Spark} 
  \label{Architecture}
\end{figure}

\subsubsection{Real-Time Video Streaming Window}
The video streaming window provides users with a live feed of the environment as perceived by the robotic agent. Empowered by the YOLO v7 Tiny object detection model \cite{wang2023yolov7}, detected objects are visually highlighted with surrounding rectangles and accompanied by corresponding labels. This real-time object recognition mechanism delivers immediate and informative visual feedback to users, enhancing their understanding of the environment. The utilization of advanced object detection technology enriches the user experience by providing a dynamic and visually engaging interface.

\subsubsection{Chat Window} \label{chat_window}
The chat window serves as the interface for user input and interaction. It supports two distinct modes, Speech Input and Text Input, catering to different age groups and preferences. The Speech Input mode, considered the most natural and intuitive option for younger children, is initiated by clicking the microphone symbol button. In this mode, the system captures the user's spoken thoughts, automatically terminating the recording after a pause of a certain duration. This design ensures smooth and efficient interaction, aligning with the cognitive needs of younger users and enhancing the overall user experience. For older children, the Text Input mode is available, allowing input through keyboard typing. This provides flexibility in communication, accommodating users who may prefer text-based interaction. The dual-input approach caters to various age groups, making the system inclusive and adaptable to different user preferences. 

\subsection{Voice Interface}
In recognition of our text-based chatbot model, primarily designed for textual input and output, a Voice User Interface (VUI) is integrated to accommodate younger users who prefer speech-formatted interactions. This integration robust support for both speech input and output. 

\subsubsection{Speech-to-Text Integration} 
To facilitate the transition from user speech to text-based input, Spark incorporates a Speech-to-Text (STT) component. Leveraging the cutting-edge WhisperX technology \cite{bain2022whisperx}, the speech recorded through the web user interface (as described in Section \ref{chat_window}) is transmitted to the WhisperX module. WhisperX excels in efficiently converting users' spoken words into corresponding transcriptions that align with the requirements of our text-based chatbot model. This integration ensures a smooth and accurate translation of spoken input into a format compatible with the system's underlying chatbot model. 

\subsubsection{Text-to-Speech Synthesis} 
Conversely, to ensure a cohesive user-system interaction, text-based outputs from the chatbot undergo transformation into synthesized speech using a Text-to-Speech (TTS) component. This process harnesses the capabilities of the Coqui Python Text-to-Speech library\footnote{https://github.com/coqui-ai/TTS}. The outcome is synthesized speech that faithfully represents the text-based responses generated by the chatbot model. The synthesized result is then transmitted to the robot by the chatbot model, and the robot articulates the synthesized speech. The synthesized speech is played through the speaker of the robotic companion, creating an immersive auditory experience for users. This auditory feedback not only enhances engagement but also provides an additional layer of accessibility, making the interaction more inclusive and accommodating for younger users.

\subsubsection{Integration with Robotic Agent}
Through the seamless integration of VUI utilizing STT and TTS modules, our system offers younger children an immersive experience reminiscent of conversing with the robot. As demonstrated in research on VUI \cite{tutletalk_vui}, VUI enhances children's engagement in programming activities. However, in contrast to research solely relying on VUI without a tangible interface, our system distinguishes itself by incorporating a robotic agent that serves as a tangible interface. This union of VUI and a tangible interface, represented by the robotic companion, is designed to provide a more engaging and enriching experience for younger children. The coupling of VUI and a tangible interface within Spark aims to synergistically enhance the interactive and hands-on nature of the programming experience. This integrated approach ensures that users not only engage in dynamic VUI interactions but also witness tangible outcomes through the actions of the robotic companion, fostering a comprehensive and captivating programming environment for younger users.

%% file: spark_chatbot_design_CJ.tex
\section{Conversation Engine}
\label{sec_chatbot}

We build the chatbot model within Spark based on the Rasa framework \cite{bocklisch2017rasa}, an open-source tool tailored for crafting customized chatbots, with a number of enhancement for interacting with younger children for learning computational thinking. The chatbot model adheres to a turn-taking conversation flow. This structured interaction model revolves around users providing natural language input and awaiting the system's response before proceeding.

\subsection{Interaction Initiation}
The conversation initiation is prompted by user greetings, such as ``hey spark'' or ``hi spark''. In response to these greetings, the chatbot delivers a default acknowledgment: ``Yes, What do you want me to do?''. This establishes the foundation for a meaningful user-system engagement. Following the default acknowledgment, the chatbot anticipates the user to provide a desired instruction, which will be translated into an executable program. However, the chatbot model is designed to be flexible. If it recognizes a different intention based on the user's input, it generates a corresponding response. For instance, if the user queries the current supported libraries with a prompt like ``What can you do?'' the chatbot model responds with the available program constructs, showcasing adaptability to diverse user inputs. For future enhancements, we envision a system where the conversation can be initiated by speaking user greetings without the need for explicitly clicking the microphone symbol button on the web user interface. This continuous voice interaction approach aligns with the convenience provided by popular voice assistants like Alexa and Siri, further enhancing the user experience in Spark. 

\begin{figure}[h]
  \centering
  \includegraphics[scale=0.5]{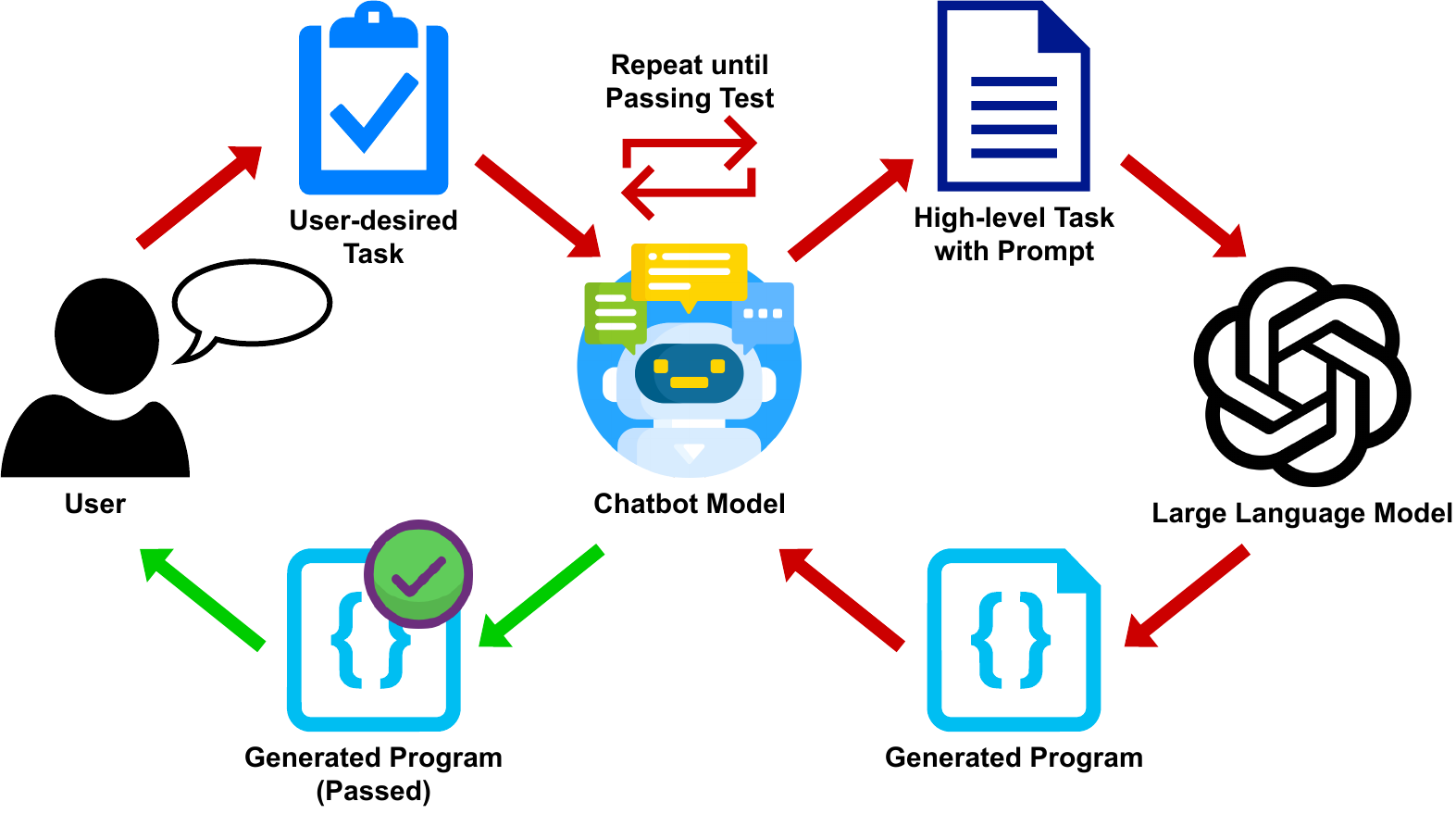}
  \caption{The generating program process diagram}
  \label{chatbot_and_llm}
\end{figure}

\subsection{Instruction Processing} \label{instruction_processing}
Upon receiving the user's desired instruction, the chatbot undertakes the task of formulating the instruction. Subsequently, it forwards this formulated instruction to the LLM along with the prepared prompt (details provided in Section \ref{LLM_for_decomposition}). This marks the initiation of the task decomposition process. The LLM, acting as a pivotal component in task decomposition, processes the received instruction. Leveraging its capabilities, the LLM transforms the complex and abstract user input into a logical sequence of executable components. This transformation is fundamental in breaking down high-level tasks into manageable and executable parts. To ensure the integrity of the generated program, the chatbot model internally validates the result obtained from the LLM (illustrated in Figure \ref{chatbot_and_llm}). If the result program fails the validation test, typically due to the existence of unsupported components or syntax errors, the chatbot model retains its internal state. In such instances, the chatbot requests the user to input a different instruction. On successful validation, the generated program is presented to the user. This program serves as a visual representation of the logical breakdown of the task. This instructional processing mechanism serves as a catalyst for younger children to grasp the concept of task decomposition from computational thinking. By breaking down high-level and complex tasks into available low-level program constructs, this approach maintains interest and fosters a foundational understanding of programming principles among younger users. 

\subsection{User Review and Decision}
Following the generation of the program with numbering, the chatbot model presents it to the user along with the query ``Do you want me to try?''-initiating the user review and decision-making phase. Upon user affirmation, signifying their approval of the generated program, the chatbot model proceeds to execute the program on the robotic agent. This streamlined execution process reflects the user's confidence in the presented program. In the event of a user denial, the chatbot model assesses the complexity of the instruction based on the number of lines in the result program. For simple instructions (single-line programs), a user denial results in a straightforward acknowledgment from the chatbot, such as ``Okay'' or ``All right'', concluding the conversation and respecting the user's decision. However, for complex instructions (multi-line programs), a denial prompts the chatbot model to inquire whether the user wants to fix the program. The question ``Do you want to fix the program?'' empowers the user to refine or modify the program based on their preferences. To empower users in modifying the decomposed result, Spark implements a user-driven revision process with the following steps. Firstly, the chatbot model displays the numbered program. The program is represented with block numbers assigned for different user intents: add/change/delete. For additions, program lines are numbered starting from 1, with numbers displayed before each line. It is important to note that the sequence number starts from 1, providing an intuitive approach for younger children. For changes or deletions, program lines are also numbered, excluding lines containing \texttt{END}. After presenting the numbered program, the chatbot model guides the user based on their intent. For deletions, the model prompts the user to specify the block number they want to remove. This straightforward process does not require additional information. For additions or changes, the model asks for an instruction to change into, in addition to the block number. 

The user-provided instruction for revision is forwarded to the LLM along with the prompt for revision (details described in Section \ref{LLM_for_decomposition}). If additional new instructions are present, the chatbot model prompts the user to decompose them, handling affirmations and denials accordingly. This dedicated process ensures users can define another new instruction recursively while defining a new one, offering an intuitive way to define the program that supports top-down computational thinking. This intuitive approach ensures a cohesive and structured program definition. The chatbot model validates and processes the result from the LLM, ensuring it aligns with the syntax and requirements outlined in Section \ref{instruction_processing}. The actual process of adding, changing, or deleting blocks is managed by a dedicated action including recognizing the user-desired block number. The chatbot model employs Duckling\footnote{https://github.com/facebook/duckling} to extract entities in various languages. With Duckling, the chatbot model obtains the user's desired block number in integer format from text-presented numbers like ``three'' or ``five''. Duckling can also extract numbers from ordinal presentations such as ``first'' and ``second''. Thus, if the user speaks their desired block number as a cardinal or ordinal number, the chatbot model recognizes the block number from the transcription. The chatbot model executes the following actions based on user intent:
\begin{itemize}
    \item Changing a Block: Modifies the program at the specified block number, handling indentation and potential changes to \texttt{END} blocks;
    \item Adding a Block: Inserts a new block at the specified location, considering the potential addition of an \texttt{END} block;
    \item Deleting a Block: Removes the specified block, managing the potential removal of associated \texttt{END} blocks.
\end{itemize}
The user-driven revision process ensures that users have control over the modification of their programs, fostering an interactive and collaborative programming environment within Spark. 

\subsection{Program Execution and Function Storage}
After a successful execution on the robotic agent, users are presented with the option to store their designed program as a reusable function. In the case of an affirmation to register a new instruction, the chatbot model proceeds to register it. The registered functions serve as sub-components for the LLM to construct the corresponding program for a given user instruction. This registration for future use provides younger children with different user-defined levels of abstraction, supporting bottom-up computational thinking. This cumulative and modular approach empowers users in constructing programs, contributing to a rich and evolving programming experience. After registration, if there are no remaining definitions due to recursive definition, the chatbot model acknowledges the registration with an ``Okay''. However, if there are additional instructions to define, the chatbot model shows the revising decomposed program and revises it with the user before confirming execution. This iterative and interactive process ensures a comprehensive user experience, allowing users to actively participate in the evolution of their program definitions. 

%% file: spark_robot_agent_CJ.tex
\section{Robotic Agent}
\label{sec_robot}

The robotic agent within our system serves as a tangible interface specifically designed for younger children, offering a real-world manifestation of their defined programs. This interaction provides a dynamic and engaging experience, bridging the virtual programming realm with tangible, physical outcomes. The utilization of the Unitree Go1 robot contributes to the enhancement of our educational platform. This integration presents a distinctive opportunity to amplify the fundamental functionalities that constitute the lowest level of programming constructs, surpassing the capabilities offered by traditional educational platforms. The tangible interaction with the robotic agent serves as a pivotal bridge between the virtual and physical dimensions, ensuring a comprehensive and holistic learning experience tailored for younger children. 

\subsection{Unitree Go1}
For illustration purpose, our current choice of the robotic agent is a quadruped robot Unitree Go1, which has a dog-like appearance to appeal to young children's familiarity. Illustrated in Figure \ref{Architecture}, the Unitree Go1 robot comprises five crucial components: the Main Control Board, the Motion Control Board, and three Sensing Motherboards. The Motion Controller Unit (MCU), acting as the Main Control Board, is responsible for the intricate control of the robot's 12 joint motors, providing it with a remarkable 12 degrees of freedom. The Motion Control Board, powered by a Raspberry Pi 4B, acts as a pivotal intermediary between developers and the internal boards. Three NVIDIA Nanos, strategically placed (one in the head, two in the body), function as the sensory components of the robot. These modules play a vital role in data acquisition and processing. The Go1 robot is equipped with a Wi-Fi module, establishing a wireless connection with the Raspberry Pi 4B, facilitating high-level command transmission to the Motion Control Board.

The Go1 robot boasts an array of sensors, including five Fish-eye Stereo Depth Cameras (front face, chin, left side of the body, right side of the body, and belly), three sets of hypersonic sensors (front face, left side of the body, and right side of the body), and a built-in speaker on its head. The front face cameras deliver real-time visual feedback to younger learners, showcasing recorded frames with highlighted objects detected by an object detection model. The ultrasound modules measure distances to surrounding objects, ensuring the robot avoids collisions. Moreover, they serve as a safety net, halting actions if the proximity to nearby objects falls below a predefined threshold, such as 30 cm. The inclusion of a built-in speaker on the Go1's head further enriches the interaction experience. This speaker plays synthesized speech transmitted from the chatbot model, contributing significantly to the auditory engagement of younger children during interactions with the robot. This comprehensive set of features ensures an engaging and safe interaction for our target audience.

\subsection{Spark SDK}
To facilitate control of the Go1 robot, a Python program has developed, harnessing the capabilities of the Software Development Kit (SDK). This program enables high-level control over the Go1 robot by transmitting motion commands, including forward, left, right, and backward movements, directly to the robot. Communication between the Python program and the Go1 robot is established through the UDP protocol, ensuring real-time data exchange for responsive control and telemetry. The SDK defines the command structure, and a thorough understanding of its syntax is imperative for precise command transmission. Commands are encoded into logical sequences and transmitted over the UDP connection. Concurrently, the robot reciprocates by sending data packets containing information about its sensors and internal state. This bidirectional communication sets the stage for a robust feedback loop, allowing the host machine to receive and parse data, thereby gaining valuable insights into the robot's status. The bidirectional communication establishes a feedback loop where incoming sensor data informs subsequent commands. This dynamic interaction ensures stable and accurate operation, enhancing the overall user experience and interaction with the robot. The sophistication embedded in this Python program plays a pivotal role as an interface for high-level control of the Go1 robot. By utilizing the SDK and the UDP protocol, it facilitates real-time communication, precise command execution, and responsive feedback loops. This program stands as a testament to the commitment to achieving optimal control and interaction with the Go1 robot within the educational context.

\subsection{Basic Actions}
In conjunction with a Python program and Spark Programming Language (SPL), detailed in Section \ref{SPL}, the Unitree Go1 robot supports a set of fundamental actions. These actions, executable through SPL, encompass various movements and sensory interactions, enhancing the versatility of the robotic agent. 

The repertoire of movement actions encompasses a diverse set of commands designed to facilitate the Unitree Go1 robot's dynamic interactions, allowing it to perform various maneuvers and motions within its environment. The movement actions include: 
\begin{itemize}
\item \texttt{STAND\_DOWN}, \texttt{STAND\_UP}: Facilitate the transition between the Go1 robot's prone and standing postures, allowing it to lower its body to the ground or rise to an upright stance.
\item \texttt{TILT\_LEFT\_SHOULDER}, \texttt{TILT\_RIGHT\_SHOULDER}, \texttt{TILT\_HEAD\_UP}, \texttt{TILT\_HEAD\_DOWN}, \texttt{TILT\_HEAD\_LEFT}, \texttt{TILT\_HEAD\_RIGHT}: Adjust the orientation of the shoulder or head in specific directions (left, right, up, or down).
\item \texttt{MOVE\_FORWARD}, \texttt{MOVE\_LEFT}, \texttt{MOVE\_RIGHT}: Propel the Go1 robot linearly in cardinal directions without altering its facing direction.
\item \texttt{TURN\_LEFT}, \texttt{TURN\_RIGHT}: Rotate the Go1 robot on its vertical axis, changing the direction it faces.
\item \texttt{SPIN\_JUMP}: Combine rotational spin with a vertical leap, enabling the robot to perform an aerial maneuver.
\item \texttt{LIFT}: Initiate a pre-programmed mimicking motion of a dog standing on its hind legs and shaking its front paws.
\item \texttt{FIRST\_DANCE, SECOND\_DANCE}: Initiate pre-programmed dance sequences, each with distinct choreography patterns.
\item \texttt{FIND}: Search for the specified object (\texttt{OBJECT}) by rotating 360 degrees slowly. If found, store the information to return \texttt{True} when queried with \texttt{FOUND OBJECT}.
\end{itemize}
All movement actions are designed to ensure safety and efficiency, being preceded by collision detection. The \texttt{FIND} action is implemented using the YOLO v7 Tiny object detection model \cite{wang2023yolov7}. The process involves feeding the recorded image frame from the front camera into the object detection model after pre-processing, which includes resizing and normalizing. The model outputs predictions for detected objects, including location, size, and confidence scores. To refine the results, only predictions with a high confidence score are retained. Non-Maximum Suppression (NMS) is applied to reduce multiple boxes for the same object to the single most accurate box. The original image is annotated with final bounding boxes and class labels to highlight detected objects. The Go1 robot continually checks for detected objects while rotating, concluding the search successfully if the desired object (\texttt{OBJECT}) is within the field of view, indicated by center coordinates between predefined thresholds (e.g., 0.4 and 0.6). Movement commands are then sent accordingly. If the desired object is not found, the search ends.

The basic sensing actions, producing True or False values for \texttt{IF} and \texttt{WHILE} in SPL, are crucial components of the Go1 robot's interactive capabilities. These actions include:
\begin{itemize}
\item \texttt{LIGHT}, \texttt{DARK}: Evaluates the lighting conditions in the environment. \texttt{LIGHT} is True if the environment is well-lit, and \texttt{DARK} is True if the environment is not lit. Otherwise, both are False;
\item \texttt{FAR}, \texttt{NEAR}: Assess the robot's proximity to objects. \texttt{FAR} is True if the robot is beyond specified distances from any object in the face, left, or right direction. \texttt{NEAR} is True if the robot is within specified distances from any object in the face, left, or right direction. Otherwise, both are False;
\item \texttt{FOUND}: Returns True, with the given argument \texttt{OBJECT} in the syntax \texttt{FOUND OBJECT}, if the specified object is found using the \texttt{FIND} action. Otherwise, it is False.
\end{itemize}

Given that the Go1 robot lacks a dedicated sensor for measuring lightness, a proxy method is employed to implement \texttt{LIGHT} and \texttt{DARK}. The process involves extracting the value of the L channel from an LAB-formatted frame, which is converted from the RGB frame recorded by the camera. The L channel values are then normalized, and True is returned if the mean of the normalized values surpasses a specified threshold; otherwise, False.

These basic sensing actions serve as fundamental program constructs for younger children and the LLM. They enable the Go1 robot to perceive and respond to environmental conditions, fostering an engaging and educational interaction with users.

\subsection{Tangible Interface}
The incorporation of a tangible interface aligns with established research advocating for the implementation of tangible interfaces customized for the educational needs of young learners \cite{kibo_robot_demo}. This strategic integration within Spark's framework involves the utilization of the Go1 robot as a pivotal component. The rationale behind this design decision is rooted in the assertion that tangible interfaces significantly augment the learning experience, particularly for a youthful demographic. The Go1 robot, boasting a comprehensive set of functionalities encompassing cameras and speakers, serves as a multifaceted and immersive learning platform. This choice of tangible interface serves a dual purpose. First, it aligns with the broader educational objective of catering to the cognitive and sensory preferences of young learners. Second, it leverages the advanced technological capabilities of the Unitree Go1 robot to provide a rich and interactive educational environment. The inclusion of cameras facilitates visual interaction, enabling users to engage with the learning content in a more visually comprehensive manner. Meanwhile, the incorporation of speakers enhances auditory engagement, contributing to a holistic learning experience. 

\subsection{Enhanced Programming Comprehension}
The integration of the Unitree Go1 robot within Spark's architecture extends beyond its tangible interface, encompassing an approach to interactive learning. Central to this approach is the utilization of synthesized speech from the robot's speaker, a feature designed to elevate programming comprehension. The synthesized speech serves a dual purpose by providing not only auditory feedback but also contributing to the creation of a highly immersive learning environment. This strategic use of voice attribution transforms the robotic agent from a mere machine into a relatable entity, fostering a sense of connection and engagement, particularly among young learners. In contrast to conventional educational platforms such as ScratchJr and KIBO, the Go1 robot's basic functionalities transcend the norm. These capabilities lay a robust foundation for interactive learning experiences that go beyond the conventional boundaries of educational programming tools. The incorporation of synthesized speech elevates the educational interaction, making the learning process more dynamic and engaging. Moreover, the inherent capabilities of the Unitree Go1 robot not only meet but exceed the standards set by traditional educational platforms. This technological advancement opens up new dimensions for educational innovation, providing a platform that goes beyond the limitations of existing tools. Beyond the current capabilities, the Go1 robot's potential for future expansion through combinations of lower-level controls introduces a forward-looking perspective. This flexibility paves the way for continuous development and innovation in the realm of interactive learning. As such, the integration of synthesized speech is not only a current enhancement but also a strategic investment in the future adaptability and growth of Spark's educational platform.

%% file: spark_spark_programming_language_CJ.tex
\section{Spark Programming Language} \label{SPL}
The paradigm of programming languages traverses a spectrum, commencing with high-level languages like Python and C, characterized by human readability and cross-platform portability. Progressing further, we encounter middle-level assembly languages such as x86 and ARM, acting as a conduit between high-level languages and machine code, offering a balance of control over hardware and legibility. Culminating at the lowest tier is low-level machine code, like x86 and ARM machine code, constituting binary instructions directly interpretable by the computer's central processing unit (CPU). This hierarchical progression allows programmers to navigate between abstraction and precision based on the specific needs of their tasks. High-level programming languages, exemplified by Python and C, facilitate a superior level of abstraction. This abstraction affords programmers the ability to articulate ideas in a human-readable and comprehensible manner. The syntax of high-level languages closely mirrors natural language, enhancing code development and comprehension. Moreover, the inherent portability of these languages across diverse platforms arises from their design independence from underlying hardware. In the intermediary stratum, we encounter assembly languages like x86 and ARM assembly languages. Positioned as middle-level languages, they serve as a link between high-level languages and machine code. Assembly language provides a symbolic representation of machine code instructions, offering improved readability compared to machine code. While tied to specific computer architectures, it grants programmers a heightened efficiency and hardware customization, necessitating a nuanced understanding of the underlying system. In the final echelon, low-level machine code, including x86 and ARM machine code, comprises binary instructions directly executable by the computer's CPU. This form of code is intricately tied to the CPU architecture and lacks portability across different processor types. It represents the lowest-level manifestation of a program that the CPU can directly execute. 

\begin{figure}[h]
  \centering
  \includegraphics[scale=0.5]{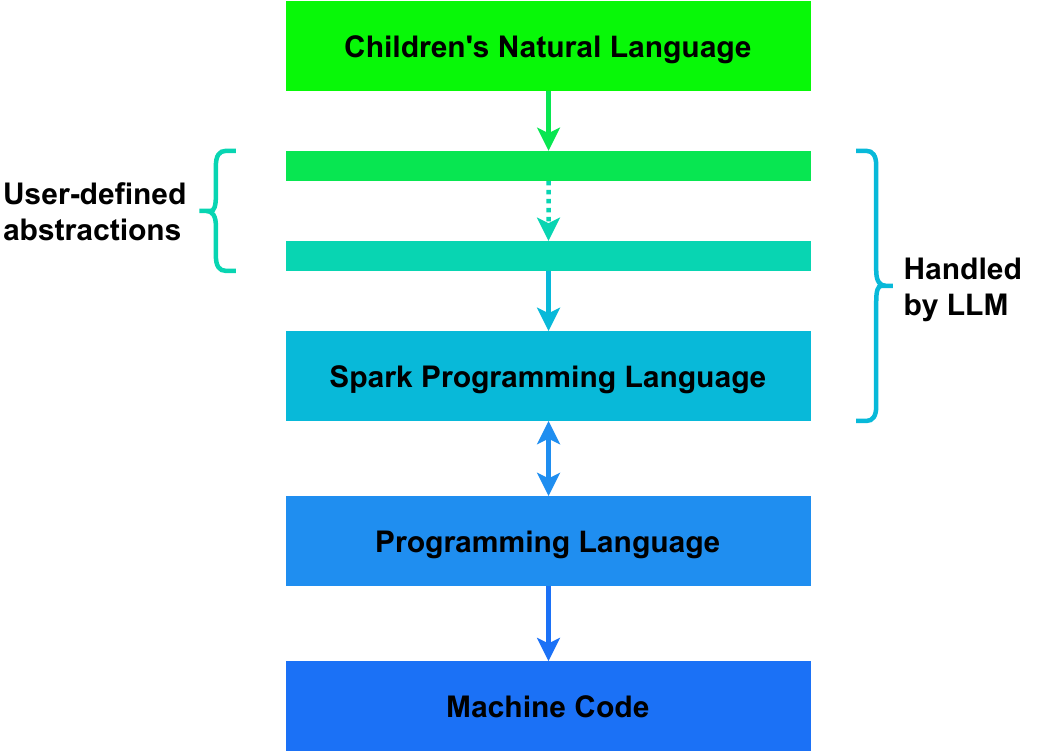}
  \caption{The abstraction hierarchy from machine code to children's natural language}
  \label{abstraction_hierarchy}
\end{figure}

Despite the advantages offered by high-level programming languages, novice learners often encounter challenges due to syntax disparities with natural languages. This hurdle is particularly pronounced for our target users, children aged 4-9, who are in the process of developing language skills, especially in reading. The conventional syntax of high-level languages may appear daunting for this age group. In response to this challenge, we introduce the Spark Programming Language (SPL), a language explicitly tailored to the needs of younger users. As illustrated in Figure \ref{abstraction_hierarchy}, SPL is designed to align its syntax more closely with natural language, thereby easing the transition into programming for our target demographic. This intentional alignment aims to mitigate the potential difficulties posed by syntax disparities, offering a more accessible entry point for young learners. Furthermore, SPL incorporates the concept of user-defined abstraction levels, effectively bridging the gap between SPL and the more abstract natural language commonly used by children. In the context of SPL, the LLM assumes responsibility for handling the libraries associated with the language. This strategic approach is aimed at simplifying the programming experience for younger users by offloading the complexity of libraries onto the LLM. By incorporating user-defined abstractions and leveraging the LLM for library management, SPL endeavors to provide a smoother learning curve for programming, especially tailored to the varying language development stages of younger users.

\subsection{Python Base}
The cornerstone of our system lies in the utilization of Python as the operational language for our robotic agent. Consequently, any code intended for execution on the agent must be composed in Python. Given that SPL serves as an intermediary language for younger children interacting with the robotic agent, the necessity arises for seamless translation of SPL programs into Python. This translational capability ensures the operability of the robotic agent while preserving a syntax that resonates closely with the language comprehension of children. The foundation of SPL is intricately intertwined with Python, not only for operational compatibility but also to provide a harmonious learning journey for children. The design considerations prioritize interchangeability and adherence to readability principles, collectively contributing to an effective and user-friendly programming interface.

\subsubsection{Code Interchangeability}
SPL, acting as an interface for children, is inherently designed to be interchangeable with Python. The symbiotic relationship between SPL and Python is essential, as it enables the translation of SPL code into Python, facilitating the execution of commands on the robotic agent. The emphasis on interchangeability underscores our commitment to providing children with a coding experience that mirrors the syntax of their familiar language.

\subsubsection{Readability Rules}
Recognizing the significance of readability in programming, particularly for younger users, SPL adopts several readability-enhancing rules inspired by Python. The prominent example is the indentation rule, a hallmark of Python's readability philosophy. By aligning SPL with Python's readability principles, we not only ensure compatibility but also create a cohesive learning environment. This strategic alignment simplifies the transition from SPL to Python, promoting a seamless and comprehensible coding experience for our young users.

\subsection{Basic Actions}
In conjunction with a Python program and SPL, the Unitree Go1 robot supports a set of fundamental actions. These actions, executable through SPL, encompass various movements and sensory interactions, enhancing the versatility of the robotic agent. 

\subsubsection{Movement Actions}
The repertoire of movement actions encompasses a diverse set of commands designed to facilitate the Unitree Go1 robot's dynamic interactions, allowing it to perform various maneuvers and motions within its environment. The movement actions include: 
\begin{itemize}
\item \texttt{STAND\_DOWN}, \texttt{STAND\_UP}: Facilitate the transition between the Go1 robot's prone and standing postures, allowing it to lower its body to the ground or rise to an upright stance.
\item \texttt{TILT\_LEFT\_SHOULDER}, \texttt{TILT\_RIGHT\_SHOULDER}, \texttt{TILT\_HEAD\_UP}, \texttt{TILT\_HEAD\_DOWN}, \texttt{TILT\_HEAD\_LEFT}, \texttt{TILT\_HEAD\_RIGHT}: Adjust the orientation of the shoulder or head in specific directions (left, right, up, or down).
\item \texttt{MOVE\_FORWARD}, \texttt{MOVE\_LEFT}, \texttt{MOVE\_RIGHT}: Propel the Go1 robot linearly in cardinal directions without altering its facing direction.
\item \texttt{TURN\_LEFT}, \texttt{TURN\_RIGHT}: Rotate the Go1 robot on its vertical axis, changing the direction it faces.
\item \texttt{SPIN\_JUMP}: Combine rotational spin with a vertical leap, enabling the robot to perform an aerial maneuver.
\item \texttt{LIFT}: Initiate a pre-programmed mimicking motion of a dog standing on its hind legs and shaking its front paws.
\item \texttt{FIRST\_DANCE, SECOND\_DANCE}: Initiate pre-programmed dance sequences, each with distinct choreography patterns.
\item \texttt{FIND}: Search for the specified object (\texttt{OBJECT}) by rotating 360 degrees slowly. If found, store the information to return \texttt{True} when queried with \texttt{FOUND OBJECT}.
\end{itemize}
All movement actions are designed to ensure safety and efficiency, being preceded by collision detection. The \texttt{FIND} action is implemented using the YOLO v7 Tiny object detection model \cite{wang2023yolov7}. The process involves feeding the recorded image frame from the front camera into the object detection model after pre-processing, which includes resizing and normalizing. The model outputs predictions for detected objects, including location, size, and confidence scores. To refine the results, only predictions with a high confidence score are retained. Non-Maximum Suppression (NMS) is applied to reduce multiple boxes for the same object to the single most accurate box. The original image is annotated with final bounding boxes and class labels to highlight detected objects. The Go1 robot continually checks for detected objects while rotating, concluding the search successfully if the desired object (\texttt{OBJECT}) is within the field of view, indicated by center coordinates between predefined thresholds (e.g., 0.4 and 0.6). Movement commands are then sent accordingly. If the desired object is not found, the search ends.

\subsubsection{Sensing Actions}
The basic sensing actions, producing True or False values for \texttt{IF} and \texttt{WHILE} in SPL, are crucial components of the Go1 robot's interactive capabilities. These actions include:
\begin{itemize}
\item \texttt{LIGHT}, \texttt{DARK}: Evaluates the lighting conditions in the environment. \texttt{LIGHT} is True if the environment is well-lit, and \texttt{DARK} is True if the environment is not lit. Otherwise, both are False;
\item \texttt{FAR}, \texttt{NEAR}: Assess the robot's proximity to objects. \texttt{FAR} is True if the robot is beyond specified distances from any object in the face, left, or right direction. \texttt{NEAR} is True if the robot is within specified distances from any object in the face, left, or right direction. Otherwise, both are False;
\item \texttt{FOUND}: Returns True, with the given argument \texttt{OBJECT} in the syntax \texttt{FOUND OBJECT}, if the specified object is found using the \texttt{FIND} action. Otherwise, it is False.
\end{itemize}

Given that the Go1 robot lacks a dedicated sensor for measuring lightness, a proxy method is employed to implement \texttt{LIGHT} and \texttt{DARK}. The process involves extracting the value of the L channel from an LAB-formatted frame, which is converted from the RGB frame recorded by the camera. The L channel values are then normalized, and True is returned if the mean of the normalized values surpasses a specified threshold; otherwise, False.

\subsection{Syntax Structure}
The syntax of Spark Programming Language (SPL) is crafted to prioritize capitalization uniformity, integrate predefined robotic actions, simplify repetitive statements, offer clear flow control expressions, and maintain a bracket-free structure for functions. These design elements collectively converge to create an accessible and user-friendly programming language, specifically tailored to meet the cognitive needs of younger users.

\subsubsection{Capitalization and Naming Conventions}
SPL adopts a consistent capitalization approach, employing uppercase letters exclusively. This intentional decision is driven by the desire to enhance code recognition and optimize the visual coherence of the programming language. Additionally, naming conventions in SPL permit only capital letters and the underscore (\texttt{\_}) for function names. This constraint is strategically enforced to ensure clarity and maintain uniformity throughout the codebase. 

\subsubsection{Built-in Functions for Robotic Actions}
Integral to the operability of the robotic agent, SPL incorporates a suite of built-in functions. These functions encompass both movement actions (e.g., \texttt{STAND\_DOWN}, \texttt{MOVE\_LEFT}) and sensory actions (e.g., \texttt{LIGHT}, \texttt{NEAR}). The inclusion of these functions aligns SPL with the practical requirements of interacting with the robotic agent, enhancing its versatility and utility.

\subsubsection{Repetitive and Flow Control Statements}
SPL introduces a simplified representation for repetitive statements using the syntax \texttt{REPEAT N TIMES}, where \texttt{N} is a positive integer specified by the user. This construct provides a more straightforward alternative to Python's for-loop syntax, enhancing the comprehensibility of code for younger users. Flow control statements, including \texttt{if} and \texttt{while} statements, are expressed as \texttt{IF CONDITION} and \texttt{WHILE CONDITION}, respectively. Each of these statements is concluded with the corresponding \texttt{END IF} or \texttt{END WHILE} statement. Additionally, \texttt{REPEAT} statements are terminated with \texttt{END REPEAT}. The code within these statements is indented with four spaces, aligning with the indentation conventions of Python. This approach contributes to a clear and visually organized structure, facilitating ease of understanding.

\subsubsection{Bracket-Free Functions}
SPL embraces a bracket-free paradigm for functions, even when they involve arguments. This design choice aims to enhance code cleanliness and simplicity. For instance, the representation of ``If a car is found'' is succinctly expressed as \texttt{IF FOUND CAR} in SPL, eliminating the need for brackets and minimizing syntactic complexity. This approach contributes to an uncluttered and streamlined syntax, facilitating a smoother learning experience for younger users.

\subsection{Conversion Process}
While SPL serves as the expressive medium for presenting programs to users and the LLM, the actual execution of these programs occurs within the system implemented in Python. To bridge this gap, SPL programs undergo conversion to equivalent Python code before being executed on the robotic agent. Despite the syntactic disparities between SPL and Python, the interchangeability of the two ensures a seamless integration into the system's execution process. It is important to note that SPL, while syntactically distinct from Python, is designed for mutual intelligibility and interchangeability within the system. The conversion process ensures that the cognitive needs of younger users are met through SPL while seamlessly integrating with the Python-based execution environment.

\subsubsection{Conversion to Python Syntax}
The conversion process involves translating SPL syntax into Python syntax, ensuring compatibility and adherence to Python's execution requirements. Our dedicated converter facilitates this translation, converting simplified SPL code to standard Python syntax. Noteworthy transformations include:
\begin{itemize}
    \item \texttt{REPEAT N TIMES} becomes \texttt{for \_ in range(N):};
    \item \texttt{IF CONDITION} transforms into \texttt{if condition():};
    \item \texttt{WHILE CONDITION} is translated to \texttt{while condition():};
    \item \texttt{FIND OBJECT} converts to \texttt{find('object')}.
\end{itemize}
This conversion ensures that SPL code, designed with younger users in mind, seamlessly integrates into the Python execution environment.

\subsubsection{Conversion to SPL Syntax}
Conversely, SPL programs can be translated from standard Python syntax back to the simplified syntax used within SPL. Our converter facilitates this process, recognizing Python constructs and converting them to their SPL equivalents. Notable transformations include:
\begin{itemize}
    \item \texttt{for \_ in range(N):} becomes \texttt{REPEAT N TIMES};
    \item \texttt{if condition():} transforms into \texttt{IF CONDITION};
    \item \texttt{while condition():} is translated to \texttt{WHILE CONDITION};
    \item \texttt{find('object')} converts to \texttt{FIND OBJECT}.
\end{itemize}
This bidirectional conversion enables the flexibility to switch between SPL and Python representations.

%% file: spark_llm_for_nlp_to_spl_CJ.tex
\section{LLM for task decomposition} \label{LLM_for_decomposition}
The pivotal component within our system is the Large Language Model (LLM), representing the core intelligence. Its primary function is to facilitate the semantic decomposition of high-level tasks into more manageable, lower-level tasks. The utilization of GPT-3.5 \cite{NEURIPS2020_1457c0d6, ye2023comprehensive} further elevates the capabilities of our chatbot model, enabling it to comprehend and navigate intricate instructions through the process of semantic decomposition. By harnessing the power of the LLM, our system provides an intuitive platform for users, especially those of a younger demographic, to witness the transformation of their desired instructions into comprehensible sub-components. This process not only enhances the chatbot's ability to understand complex commands but also contributes to the development of top-down computational thinking among users. 

\begin{figure}[h]
  \centering
\begin{mybox}{}
\begin{minipage}[t]{0.6\linewidth}
\texttt{Think step by step to carry out the instruction.\\
Write an executable program for the robot based on given instructions.\\
Use only available options to write the program.\\
Try to use the new available options as much as possible.\\
Ambiguous instructions must be interpreted in the most reasonable way and the program must be written accordingly.\\
\\
Available Options:\\
\{available\_options\}\\
\\
The following descriptions are for new available options.\{new\_available\_option\_pairs\}\\
\\
STAND\_UP, ..., SECOND\_DANCE take no arguments.\\
FIND only takes one argument: a string to be matched with the object name. For example, FIND CUP will find a cup. You must not use FIND with parentheses, for example, FIND(CUP) is not allowed.\\
Available objects for FIND are only as follows: \\
person\\
...\\
For repeated actions, use REPEAT N TIMES. For example, REPEAT 3 TIMES will repeat the following actions 3 times.\\
The number of times, N, must be a positive integer.\\
The repeated actions must be indented by 4 spaces.\\
After the repeated actions, write END REPEAT.\\
\\
For conditional execution, use IF CONDITION syntax. \\
For example, IF LIGHT will execute the following actions if the environment is bright.\\
The conditional actions must be indented by 4 spaces.\\
After the conditional actions, write END IF.\\
Available conditions for IF only include LIGHT, DARK, FAR, NEAR, FOUND.\\
FOUND is used along with FIND. \\
For example, IF FOUND CUP will execute the following actions if the robot found a cup with FIND CUP.\\
\\}
\end{minipage} 
\hfill\begin{minipage}[t]{0.35\linewidth}
\texttt{For conditional repetition, use WHILE CONDITION syntax. For example, WHILE FAR will repeat the following actions while the robot is far from the object.\\
...\\
Instructions can be varied in format, but should all be translated to the same program if they have the same meaning.\\
Examples:\\
\{\{instruction\}\}\\
I want to \{\{instruction\}\}\\
...\\
For each form of instruction received, translate it into the robot's programming language.\\
Instruction: go\\
Program:\\
MOVE\_FORWARD\\
...\\
Instruction: if far, then move twice\\
Program:\\
IF FAR\\
{\transparent{0}~}{\transparent{0}~}{\transparent{0}~}{\transparent{0}~}REPEAT 2 TIMES\\
{\transparent{0}~}{\transparent{0}~}{\transparent{0}~}{\transparent{0}~}{\transparent{0}~}{\transparent{0}~}{\transparent{0}~}{\transparent{0}~}MOVE\_FORWARD\\
{\transparent{0}~}{\transparent{0}~}{\transparent{0}~}{\transparent{0}~}END REPEAT\\
END IF\\
...\\
Instruction: go to the chair\\
Program:\\
FIND CHAIR\\
IF FOUND CHAIR\\
{\transparent{0}~}{\transparent{0}~}{\transparent{0}~}{\transparent{0}~}WHILE FAR\\
{\transparent{0}~}{\transparent{0}~}{\transparent{0}~}{\transparent{0}~}{\transparent{0}~}{\transparent{0}~}{\transparent{0}~}{\transparent{0}~}MOVE\_FORWARD\\
{\transparent{0}~}{\transparent{0}~}{\transparent{0}~}{\transparent{0}~}END WHILE\\
END IF\\
\{new\_instruction\_program\_pairs\}\\
Instruction: \{instruction\}\\
Program:}
\end{minipage}
\end{mybox}
  \caption{The prompt for decomposition}
  \label{prompt}
\end{figure}

\subsection{Prompt}
Our system adopts a designed prompt inspired by the methodologies presented in \cite{cvpr2023_bp, NEURIPS2022_9d560961}. This prompt serves as the guiding instruction for the LLM, directing it to generate executable programs tailored for our system's robotic agent based on provided instructions. Illustrated in Figure \ref{prompt}, the curated prompt intricately instructs the LLM on the utilization of available options, encompassing pre-defined basic actions and user-defined programs. The objective is to construct programs that exclusively employ the provided options, with an encouragement for the model to integrate new available options whenever possible. To ensure clarity and precision, the prompt addresses the interpretation of ambiguous instructions, emphasizing the imperative need for the generated program to align with the most reasonable interpretation. The available options encompass pre-defined basic actions, such as \texttt{STAND\_UP} and \texttt{MOVE\_FORWARD}, with varying argument requirements. For instance, actions like \texttt{FIND} require a single argument in the form of a string matched with an object name, drawn from the COCO 2017 Object Detection dataset\footnote{https://cocodataset.org/\#detection-2017} used for training the object detection model. Introducing conditional elements like \texttt{LIGHT}, \texttt{DARK}, \texttt{FAR}, \texttt{NEAR}, and \texttt{FOUND}, the prompt facilitates conditional execution and repetition within the generated programs. The syntax includes \texttt{REPEAT TIMES} for repeating actions and \texttt{IF CONDITION} for conditional execution, while \texttt{WHILE CONDITION} is employed for conditional repetition, utilizing conditions like \texttt{LIGHT}, \texttt{DARK}, \texttt{FAR}, and \texttt{NEAR}. Furthermore, the prompt provides illustrative examples showcasing the translation of diverse instructions into the robot's programming language. The examples span various instructions, encompassing movement, object detection, and the handling of conditional statements and repetitions. In the continuous evolution of the system, the prompt maintains an open-ended segment, allowing the LLM to receive specific instructions and generate corresponding programs in our system's programming language explained in Section \ref{SPL}. This adaptive design ensures the system's responsiveness to novel instruction-program pairs, promoting continuous learning and enhancement of its programming capabilities. 

\begin{figure}[h]
  \centering
  \includegraphics[scale=0.4]{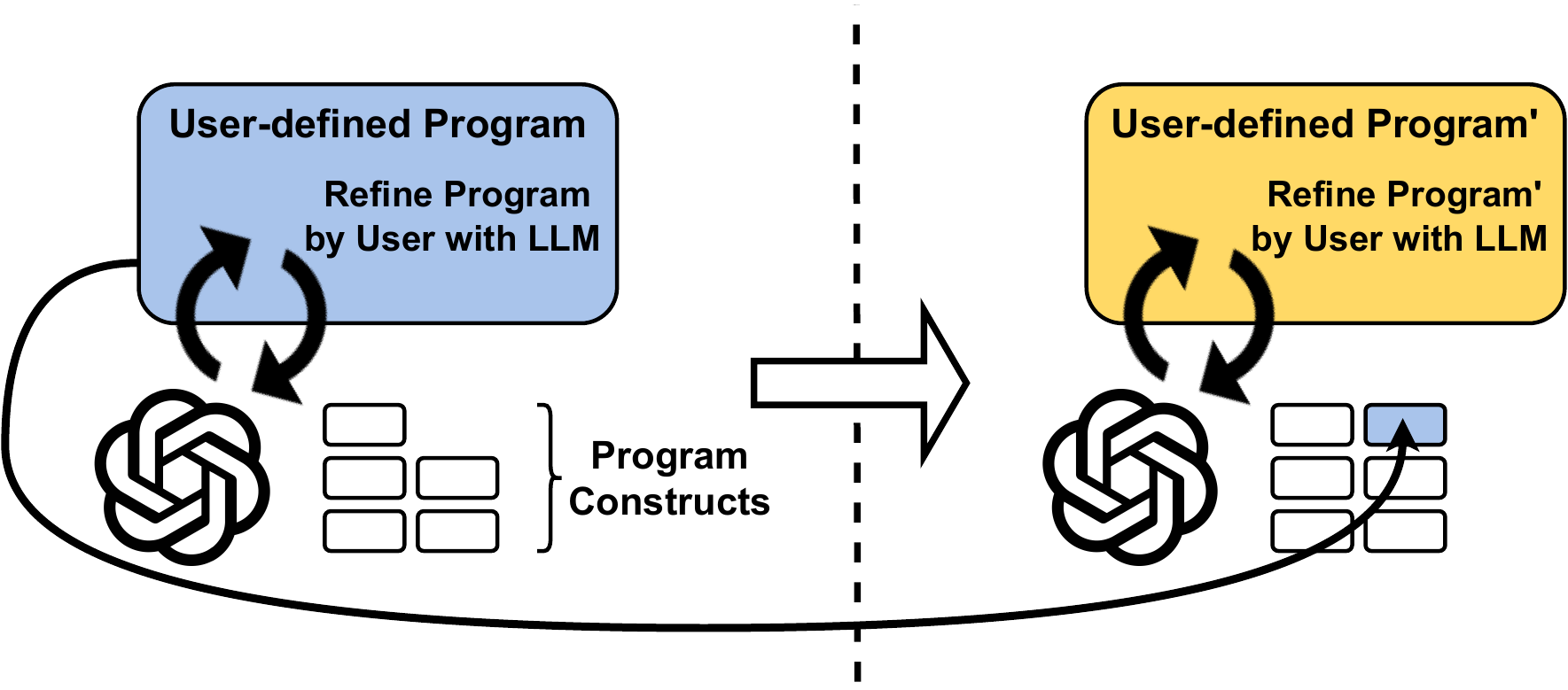}
  \caption{The user-defined programs utilization diagram}
  \label{spark_reuse}
\end{figure}

\subsection{Semantic Decomposition}
At the nucleus of our system's task decomposition lies the influential integration of GPT-3.5, a robust language model hosted by OpenAI. This strategic integration serves as a catalyst in empowering users, with a particular focus on younger children, by simplifying the intricate process of breaking down complex tasks into more manageable components. The synergy between GPT-3.5 and the crafted prompt is indispensable for the effective execution of task decomposition. Illustrated in Figure \ref{spark_reuse}, the pivotal role of user-defined functions in the task decomposition process is prominently evident. These functions merge into the prompt, assuming the crucial status of fundamental components within the system architecture. Serving as building blocks, these user-defined functions are adeptly employed by the LLM to systematically formulate programs in alignment with user instructions. Users actively participate in this process by articulating their desired functions, thereby initiating a collaborative endeavor where these functions integrate into the prompt. This user-centric methodology establishes a streamlined trajectory for task decomposition, ensuring that the user-defined functions play a substantial and influential role in the meticulous construction of programs orchestrated by the LLM. A notable attribute of our system lies in its adaptability to users with diverse skill levels, with a specific focus on addressing the needs of younger children. The initiation of programming is marked by flexibility, allowing users to commence at any level of task abstraction. This flexibility spans from the highest level of defining the final task down to the granular level of specifying functions provided by the robotic agent. Such inclusivity creates a supportive environment, empowering users to engage with programming at their individually preferred level of complexity. This approach reflects a commitment to accommodating a wide spectrum of user competencies, ensuring a comprehensive and tailored learning experience. 

\subsection{Learning Experience}
The incorporation of the LLM in task decomposition holds the potential to significantly enhance the learning experience, particularly for younger children. Our strategic use of the LLM focuses on empowering these young users to intuitively grasp the process of breaking down intricate tasks into more manageable, lower-level components. The core objective is to cultivate a learning environment where younger children can develop a profound understanding of task decomposition. Leveraging the LLM's capabilities, users are guided through the systematic deconstruction of their high-level objectives into accessible, comprehensible components. This approach not only facilitates a more profound comprehension of task decomposition but also addresses the challenge of sustaining interest among younger users. In contrast to manual decomposition or the construction of basic components, which may require extended durations and risk inducing boredom, our system, backed by the LLM, ensures a dynamic and engaging learning experience. Younger users benefit from a process that is not only educational but also captivating, fostering a sense of curiosity and exploration.

%% file: conclusion_CJ.tex
\section{Conclusion and Future Work}
\label{sec_conclusion}

In this study, we present Spark, an innovative system designed to cultivate computational thinking skills in early childhood coding education. At its core, Spark leverages a conversational robotic agent driven by a Large Language Model (LLM), which plays a pivotal role in instilling a natural grasp of coding concepts among young learners. The multifaceted system integrates a user-friendly chatbot interface, facilitating voice-based interactions. Augmenting this interface are Speech-to-Text (STT) and Text-to-Speech (TTS) modules, enabling speech-based communication. Additionally, the inclusion of a tangible robot enhances the pedagogical experience by providing real-world demonstrations of program execution, fostering an interactive and immersive learning environment.

A distinctive aspect of this study is its adoption of a top-down instructional approach, empowering children devoid of prior programming knowledge to articulate instructions. The LLM, a key enabler in this approach, adeptly translates these instructions into coherent programs, systematically breaking them down into predefined basic actions. The tangible robotic component further elevates the educational experience by offering tangible demonstrations, thereby solidifying the understanding of fundamental programming concepts.

In the realm of future work, we posit the integration of a participatory user study to elicit direct feedback from key stakeholders, including educators, parents, and the young users themselves. This participatory approach is envisioned to serve as a robust mechanism for refining Spark based on nuanced user preferences and evolving educational needs. Furthermore, the exploration of adaptive learning features within Spark emerges as a promising avenue. By tailoring the system to individual learning styles and preferences, a profound optimization of the educational impact can be achieved, thereby contributing to the continuous enhancement and evolution of Spark as an effective tool in early childhood coding education.

%% file: sample-manuscript.bbl

\begin{thebibliography}{24}


\ifx \showCODEN    \undefined \def \showCODEN     #1{\unskip}     \fi
\ifx \showDOI      \undefined \def \showDOI       #1{#1}\fi
\ifx \showISBNx    \undefined \def \showISBNx     #1{\unskip}     \fi
\ifx \showISBNxiii \undefined \def \showISBNxiii  #1{\unskip}     \fi
\ifx \showISSN     \undefined \def \showISSN      #1{\unskip}     \fi
\ifx \showLCCN     \undefined \def \showLCCN      #1{\unskip}     \fi
\ifx \shownote     \undefined \def \shownote      #1{#1}          \fi
\ifx \showarticletitle \undefined \def \showarticletitle #1{#1}   \fi
\ifx \showURL      \undefined \def \showURL       {\relax}        \fi
\providecommand\bibfield[2]{#2}
\providecommand\bibinfo[2]{#2}
\providecommand\natexlab[1]{#1}
\providecommand\showeprint[2][]{arXiv:#2}

\bibitem[Bain et~al\mbox{.}(2023)]%
        {bain2022whisperx}
\bibfield{author}{\bibinfo{person}{Max Bain}, \bibinfo{person}{Jaesung Huh}, \bibinfo{person}{Tengda Han}, {and} \bibinfo{person}{Andrew Zisserman}.} \bibinfo{year}{2023}\natexlab{}.
\newblock \showarticletitle{WhisperX: Time-Accurate Speech Transcription of Long-Form Audio}.
\newblock \bibinfo{journal}{\emph{INTERSPEECH 2023}} (\bibinfo{year}{2023}).
\newblock


\bibitem[Beals and Bers(2006)]%
        {beals2006robotic}
\bibfield{author}{\bibinfo{person}{Laura Beals} {and} \bibinfo{person}{Marina Bers}.} \bibinfo{year}{2006}\natexlab{}.
\newblock \showarticletitle{Robotic technologies: when parents put their learning ahead of their child’s}.
\newblock \bibinfo{journal}{\emph{Journal of Interactive Learning Research}} \bibinfo{volume}{17}, \bibinfo{number}{4} (\bibinfo{year}{2006}), \bibinfo{pages}{341--366}.
\newblock


\bibitem[Bers et~al\mbox{.}(2014)]%
        {BERS2014145}
\bibfield{author}{\bibinfo{person}{Marina~Umaschi Bers}, \bibinfo{person}{Louise Flannery}, \bibinfo{person}{Elizabeth~R. Kazakoff}, {and} \bibinfo{person}{Amanda Sullivan}.} \bibinfo{year}{2014}\natexlab{}.
\newblock \showarticletitle{Computational thinking and tinkering: Exploration of an early childhood robotics curriculum}.
\newblock \bibinfo{journal}{\emph{Computers and Education}}  \bibinfo{volume}{72} (\bibinfo{year}{2014}), \bibinfo{pages}{145--157}.
\newblock
\showISSN{0360-1315}
\urldef\tempurl%
\url{https://doi.org/10.1016/j.compedu.2013.10.020}
\showDOI{\tempurl}


\bibitem[Blikstein and Moghadam(2018)]%
        {blikstein2018pre}
\bibfield{author}{\bibinfo{person}{Paulo Blikstein} {and} \bibinfo{person}{Sepi~Hejazi Moghadam}.} \bibinfo{year}{2018}\natexlab{}.
\newblock \showarticletitle{Pre-college computer science education: A survey of the field}.
\newblock  (\bibinfo{year}{2018}).
\newblock


\bibitem[Bocklisch et~al\mbox{.}(2017)]%
        {bocklisch2017rasa}
\bibfield{author}{\bibinfo{person}{Tom Bocklisch}, \bibinfo{person}{Joey Faulkner}, \bibinfo{person}{Nick Pawlowski}, {and} \bibinfo{person}{Alan Nichol}.} \bibinfo{year}{2017}\natexlab{}.
\newblock \bibinfo{title}{Rasa: Open Source Language Understanding and Dialogue Management}.
\newblock
\newblock
\showeprint[arxiv]{1712.05181}~[cs.CL]


\bibitem[Brown et~al\mbox{.}(2020)]%
        {NEURIPS2020_1457c0d6}
\bibfield{author}{\bibinfo{person}{Tom Brown}, \bibinfo{person}{Benjamin Mann}, \bibinfo{person}{Nick Ryder}, \bibinfo{person}{Melanie Subbiah}, \bibinfo{person}{Jared~D Kaplan}, \bibinfo{person}{Prafulla Dhariwal}, \bibinfo{person}{Arvind Neelakantan}, \bibinfo{person}{Pranav Shyam}, \bibinfo{person}{Girish Sastry}, \bibinfo{person}{Amanda Askell}, \bibinfo{person}{Sandhini Agarwal}, \bibinfo{person}{Ariel Herbert-Voss}, \bibinfo{person}{Gretchen Krueger}, \bibinfo{person}{Tom Henighan}, \bibinfo{person}{Rewon Child}, \bibinfo{person}{Aditya Ramesh}, \bibinfo{person}{Daniel Ziegler}, \bibinfo{person}{Jeffrey Wu}, \bibinfo{person}{Clemens Winter}, \bibinfo{person}{Chris Hesse}, \bibinfo{person}{Mark Chen}, \bibinfo{person}{Eric Sigler}, \bibinfo{person}{Mateusz Litwin}, \bibinfo{person}{Scott Gray}, \bibinfo{person}{Benjamin Chess}, \bibinfo{person}{Jack Clark}, \bibinfo{person}{Christopher Berner}, \bibinfo{person}{Sam McCandlish}, \bibinfo{person}{Alec Radford}, \bibinfo{person}{Ilya Sutskever}, {and}
  \bibinfo{person}{Dario Amodei}.} \bibinfo{year}{2020}\natexlab{}.
\newblock \showarticletitle{Language Models are Few-Shot Learners}. In \bibinfo{booktitle}{\emph{Advances in Neural Information Processing Systems}}, \bibfield{editor}{\bibinfo{person}{H.~Larochelle}, \bibinfo{person}{M.~Ranzato}, \bibinfo{person}{R.~Hadsell}, \bibinfo{person}{M.F. Balcan}, {and} \bibinfo{person}{H.~Lin}} (Eds.), Vol.~\bibinfo{volume}{33}. \bibinfo{publisher}{Curran Associates, Inc.}, \bibinfo{pages}{1877--1901}.
\newblock
\urldef\tempurl%
\url{https://proceedings.neurips.cc/paper_files/paper/2020/file/1457c0d6bfcb4967418bfb8ac142f64a-Paper.pdf}
\showURL{%
\tempurl}


\bibitem[Flannery et~al\mbox{.}(2013)]%
        {designing_scratchjr}
\bibfield{author}{\bibinfo{person}{Louise~P. Flannery}, \bibinfo{person}{Brian Silverman}, \bibinfo{person}{Elizabeth~R. Kazakoff}, \bibinfo{person}{Marina~Umaschi Bers}, \bibinfo{person}{Paula Bont\'{a}}, {and} \bibinfo{person}{Mitchel Resnick}.} \bibinfo{year}{2013}\natexlab{}.
\newblock \showarticletitle{Designing ScratchJr: Support for Early Childhood Learning through Computer Programming}. In \bibinfo{booktitle}{\emph{Proceedings of the 12th International Conference on Interaction Design and Children}} (New York, New York, USA) \emph{(\bibinfo{series}{IDC '13})}. \bibinfo{publisher}{Association for Computing Machinery}, \bibinfo{address}{New York, NY, USA}, \bibinfo{pages}{1–10}.
\newblock
\showISBNx{9781450319188}
\urldef\tempurl%
\url{https://doi.org/10.1145/2485760.2485785}
\showDOI{\tempurl}


\bibitem[Gupta and Kembhavi(2023)]%
        {cvpr2023_bp}
\bibfield{author}{\bibinfo{person}{T. Gupta} {and} \bibinfo{person}{A. Kembhavi}.} \bibinfo{year}{2023}\natexlab{}.
\newblock \showarticletitle{Visual Programming: Compositional visual reasoning without training}. In \bibinfo{booktitle}{\emph{2023 IEEE/CVF Conference on Computer Vision and Pattern Recognition (CVPR)}}. \bibinfo{publisher}{IEEE Computer Society}, \bibinfo{address}{Los Alamitos, CA, USA}, \bibinfo{pages}{14953--14962}.
\newblock
\urldef\tempurl%
\url{https://doi.org/10.1109/CVPR52729.2023.01436}
\showDOI{\tempurl}


\bibitem[Initiative(2023)]%
        {ccss}
\bibfield{author}{\bibinfo{person}{Common Core State~Standards Initiative}.} \bibinfo{year}{2023}\natexlab{}.
\newblock \bibinfo{title}{Common Core State Standards for English Language Arts \& Literacy in History/Social Studies, Science, and Technical Subjects}.
\newblock
\newblock
\urldef\tempurl%
\url{https://corestandards.org/wp-content/uploads/2023/09/ELA_Standards1.pdf}
\showURL{%
\tempurl}


\bibitem[Jung et~al\mbox{.}(2019)]%
        {tutletalk_vui}
\bibfield{author}{\bibinfo{person}{Hyunhoon Jung}, \bibinfo{person}{Hee~Jae Kim}, \bibinfo{person}{Seongeun So}, \bibinfo{person}{Jinjoong Kim}, {and} \bibinfo{person}{Changhoon Oh}.} \bibinfo{year}{2019}\natexlab{}.
\newblock \showarticletitle{TurtleTalk: An Educational Programming Game for Children with Voice User Interface}. In \bibinfo{booktitle}{\emph{Extended Abstracts of the 2019 CHI Conference on Human Factors in Computing Systems}} (Glasgow, Scotland Uk) \emph{(\bibinfo{series}{CHI EA '19})}. \bibinfo{publisher}{Association for Computing Machinery}, \bibinfo{address}{New York, NY, USA}, \bibinfo{pages}{1–6}.
\newblock
\showISBNx{9781450359719}
\urldef\tempurl%
\url{https://doi.org/10.1145/3290607.3312773}
\showDOI{\tempurl}


\bibitem[Lee et~al\mbox{.}(2023)]%
        {Lee2023-xw}
\bibfield{author}{\bibinfo{person}{Joohi Lee}, \bibinfo{person}{Candace Joswick}, {and} \bibinfo{person}{Kathryn Pole}.} \bibinfo{year}{2023}\natexlab{}.
\newblock \showarticletitle{Classroom play and activities to support computational thinking development in early childhood}.
\newblock \bibinfo{journal}{\emph{Early Child. Educ. J.}} \bibinfo{volume}{51}, \bibinfo{number}{3} (\bibinfo{date}{March} \bibinfo{year}{2023}), \bibinfo{pages}{457--468}.
\newblock


\bibitem[Papadakis(2020)]%
        {Papadakis2020gy}
\bibfield{author}{\bibinfo{person}{Stamatios Papadakis}.} \bibinfo{year}{2020}\natexlab{}.
\newblock \bibinfo{title}{Robots and Robotics Kits for Early Childhood and First School Age}.  (\bibinfo{date}{11} \bibinfo{year}{2020}).
\newblock
\urldef\tempurl%
\url{https://www.learntechlib.org/p/218338}
\showURL{%
\tempurl}


\bibitem[Price and Price-Mohr(2018)]%
        {price2018evaluation}
\bibfield{author}{\bibinfo{person}{CB Price} {and} \bibinfo{person}{RM Price-Mohr}.} \bibinfo{year}{2018}\natexlab{}.
\newblock \showarticletitle{An evaluation of primary school children coding using a text-based language (Java)}.
\newblock \bibinfo{journal}{\emph{Computers in the Schools}} \bibinfo{volume}{35}, \bibinfo{number}{4} (\bibinfo{year}{2018}), \bibinfo{pages}{284--301}.
\newblock


\bibitem[Proctor(2023)]%
        {proctor2023computational}
\bibfield{author}{\bibinfo{person}{Chris Proctor}.} \bibinfo{year}{2023}\natexlab{}.
\newblock \showarticletitle{Computational thinking}.
\newblock  (\bibinfo{year}{2023}).
\newblock


\bibitem[Ragusa and Leung(2023)]%
        {s23239335}
\bibfield{author}{\bibinfo{person}{Gisele Ragusa} {and} \bibinfo{person}{Lilian Leung}.} \bibinfo{year}{2023}\natexlab{}.
\newblock \showarticletitle{The Impact of Early Robotics Education on Students’ Understanding of Coding, Robotics Design, and Interest in Computing Careers}.
\newblock \bibinfo{journal}{\emph{Sensors}} \bibinfo{volume}{23}, \bibinfo{number}{23} (\bibinfo{year}{2023}).
\newblock
\showISSN{1424-8220}
\urldef\tempurl%
\url{https://doi.org/10.3390/s23239335}
\showDOI{\tempurl}


\bibitem[Resnick et~al\mbox{.}(2009)]%
        {resnick2009scratch}
\bibfield{author}{\bibinfo{person}{Mitchel Resnick}, \bibinfo{person}{John Maloney}, \bibinfo{person}{Andr{\'e}s Monroy-Hern{\'a}ndez}, \bibinfo{person}{Natalie Rusk}, \bibinfo{person}{Evelyn Eastmond}, \bibinfo{person}{Karen Brennan}, \bibinfo{person}{Amon Millner}, \bibinfo{person}{Eric Rosenbaum}, \bibinfo{person}{Jay Silver}, \bibinfo{person}{Brian Silverman}, {et~al\mbox{.}}} \bibinfo{year}{2009}\natexlab{}.
\newblock \showarticletitle{Scratch: programming for all}.
\newblock \bibinfo{journal}{\emph{Commun. ACM}} \bibinfo{volume}{52}, \bibinfo{number}{11} (\bibinfo{year}{2009}), \bibinfo{pages}{60--67}.
\newblock


\bibitem[Stamatios(2024)]%
        {scratchjr_sys_lit_review}
\bibfield{author}{\bibinfo{person}{Papadakis Stamatios}.} \bibinfo{year}{2024}\natexlab{}.
\newblock \showarticletitle{Can Preschoolers Learn Computational Thinking and Coding Skills with ScratchJr? A Systematic Literature Review}.
\newblock \bibinfo{journal}{\emph{International Journal of Educational Reform}} \bibinfo{volume}{33}, \bibinfo{number}{1} (\bibinfo{year}{2024}), \bibinfo{pages}{28--61}.
\newblock
\urldef\tempurl%
\url{https://doi.org/10.1177/10567879221076077}
\showDOI{\tempurl}
\showeprint{https://doi.org/10.1177/10567879221076077}


\bibitem[Su and Yang(2023)]%
        {SU2023100122}
\bibfield{author}{\bibinfo{person}{Jiahong Su} {and} \bibinfo{person}{Weipeng Yang}.} \bibinfo{year}{2023}\natexlab{}.
\newblock \showarticletitle{A systematic review of integrating computational thinking in early childhood education}.
\newblock \bibinfo{journal}{\emph{Computers and Education Open}}  \bibinfo{volume}{4} (\bibinfo{year}{2023}), \bibinfo{pages}{100122}.
\newblock
\showISSN{2666-5573}
\urldef\tempurl%
\url{https://doi.org/10.1016/j.caeo.2023.100122}
\showDOI{\tempurl}


\bibitem[Sullivan et~al\mbox{.}(2015)]%
        {kibo_robot_demo}
\bibfield{author}{\bibinfo{person}{Amanda Sullivan}, \bibinfo{person}{Mollie Elkin}, {and} \bibinfo{person}{Marina~Umaschi Bers}.} \bibinfo{year}{2015}\natexlab{}.
\newblock \showarticletitle{KIBO Robot Demo: Engaging Young Children in Programming and Engineering}. In \bibinfo{booktitle}{\emph{Proceedings of the 14th International Conference on Interaction Design and Children}} (Boston, Massachusetts) \emph{(\bibinfo{series}{IDC '15})}. \bibinfo{publisher}{Association for Computing Machinery}, \bibinfo{address}{New York, NY, USA}, \bibinfo{pages}{418–421}.
\newblock
\showISBNx{9781450335904}
\urldef\tempurl%
\url{https://doi.org/10.1145/2771839.2771868}
\showDOI{\tempurl}


\bibitem[Wang et~al\mbox{.}(2023)]%
        {wang2023yolov7}
\bibfield{author}{\bibinfo{person}{Chien-Yao Wang}, \bibinfo{person}{Alexey Bochkovskiy}, {and} \bibinfo{person}{Hong-Yuan~Mark Liao}.} \bibinfo{year}{2023}\natexlab{}.
\newblock \showarticletitle{{YOLOv7}: Trainable bag-of-freebies sets new state-of-the-art for real-time object detectors}. In \bibinfo{booktitle}{\emph{Proceedings of the IEEE/CVF Conference on Computer Vision and Pattern Recognition (CVPR)}}.
\newblock


\bibitem[Wei et~al\mbox{.}(2022)]%
        {NEURIPS2022_9d560961}
\bibfield{author}{\bibinfo{person}{Jason Wei}, \bibinfo{person}{Xuezhi Wang}, \bibinfo{person}{Dale Schuurmans}, \bibinfo{person}{Maarten Bosma}, \bibinfo{person}{brian ichter}, \bibinfo{person}{Fei Xia}, \bibinfo{person}{Ed Chi}, \bibinfo{person}{Quoc~V Le}, {and} \bibinfo{person}{Denny Zhou}.} \bibinfo{year}{2022}\natexlab{}.
\newblock \showarticletitle{Chain-of-Thought Prompting Elicits Reasoning in Large Language Models}. In \bibinfo{booktitle}{\emph{Advances in Neural Information Processing Systems}}, \bibfield{editor}{\bibinfo{person}{S.~Koyejo}, \bibinfo{person}{S.~Mohamed}, \bibinfo{person}{A.~Agarwal}, \bibinfo{person}{D.~Belgrave}, \bibinfo{person}{K.~Cho}, {and} \bibinfo{person}{A.~Oh}} (Eds.), Vol.~\bibinfo{volume}{35}. \bibinfo{publisher}{Curran Associates, Inc.}, \bibinfo{pages}{24824--24837}.
\newblock
\urldef\tempurl%
\url{https://proceedings.neurips.cc/paper_files/paper/2022/file/9d5609613524ecf4f15af0f7b31abca4-Paper-Conference.pdf}
\showURL{%
\tempurl}


\bibitem[Wing(2006)]%
        {wing2006computational}
\bibfield{author}{\bibinfo{person}{Jeannette~M Wing}.} \bibinfo{year}{2006}\natexlab{}.
\newblock \showarticletitle{Computational thinking}.
\newblock \bibinfo{journal}{\emph{Commun. ACM}} \bibinfo{volume}{49}, \bibinfo{number}{3} (\bibinfo{year}{2006}), \bibinfo{pages}{33--35}.
\newblock


\bibitem[Ye et~al\mbox{.}(2023)]%
        {ye2023comprehensive}
\bibfield{author}{\bibinfo{person}{Junjie Ye}, \bibinfo{person}{Xuanting Chen}, \bibinfo{person}{Nuo Xu}, \bibinfo{person}{Can Zu}, \bibinfo{person}{Zekai Shao}, \bibinfo{person}{Shichun Liu}, \bibinfo{person}{Yuhan Cui}, \bibinfo{person}{Zeyang Zhou}, \bibinfo{person}{Chao Gong}, \bibinfo{person}{Yang Shen}, \bibinfo{person}{Jie Zhou}, \bibinfo{person}{Siming Chen}, \bibinfo{person}{Tao Gui}, \bibinfo{person}{Qi Zhang}, {and} \bibinfo{person}{Xuanjing Huang}.} \bibinfo{year}{2023}\natexlab{}.
\newblock \bibinfo{title}{A Comprehensive Capability Analysis of GPT-3 and GPT-3.5 Series Models}.
\newblock
\newblock
\showeprint[arxiv]{2303.10420}~[cs.CL]


\bibitem[Yu and Roque(2018)]%
        {survey_of_comp_kits}
\bibfield{author}{\bibinfo{person}{Junnan Yu} {and} \bibinfo{person}{Ricarose Roque}.} \bibinfo{year}{2018}\natexlab{}.
\newblock \showarticletitle{A Survey of Computational Kits for Young Children}. In \bibinfo{booktitle}{\emph{Proceedings of the 17th ACM Conference on Interaction Design and Children}} (Trondheim, Norway) \emph{(\bibinfo{series}{IDC '18})}. \bibinfo{publisher}{Association for Computing Machinery}, \bibinfo{address}{New York, NY, USA}, \bibinfo{pages}{289–299}.
\newblock
\showISBNx{9781450351522}
\urldef\tempurl%
\url{https://doi.org/10.1145/3202185.3202738}
\showDOI{\tempurl}


\end{thebibliography}
